\documentclass[numberedappendix]{emulateapj}
\usepackage[footnotes]{trackchanges}
\usepackage{natbib}
\usepackage{color}

\begin{document}

\title{THE CLUSTERING OF LUMINOUS RED GALAXIES AT z $\sim$ 0.7 FROM EBOSS and BOSS DATA}

\author{
Zhongxu~Zhai\altaffilmark{1},
Jeremy~L.~Tinker\altaffilmark{1},
ChangHoon~Hahn\altaffilmark{1},
Hee-Jong~Seo\altaffilmark{2},
Michael~R.~Blanton\altaffilmark{1},
Rita~Tojeiro\altaffilmark{3},
Hugo~O.~Camacho\altaffilmark{4,5},
Marcos~Lima\altaffilmark{4,5},
Aurelio~Carnero~Rosell\altaffilmark{6,5},
Flavia~Sobreira\altaffilmark{7,5,8},
Luiz~N.~da~Costa\altaffilmark{5,6},
Julian~E.~Bautista\altaffilmark{9},
Joel~R.~Brownstein\altaffilmark{9},
Johan~Comparat\altaffilmark{10,11},
Kyle~Dawson\altaffilmark{9},
Jeffrey~A.~Newman\altaffilmark{12},
Abhishek~Prakash\altaffilmark{12},
Alexandre~Roman-Lopes\altaffilmark{13},
Donald~P. Schneider\altaffilmark{14, 15}
}
\altaffiltext{1}{Center for Cosmology and Particle Physics, Department of Physics, New York University, 4 Washington Place, New York, NY 10003, USA.}
\altaffiltext{2}{Department of Physics and Astronomy, Ohio University, 251B Clippinger Labs, Athens, OH 45701.}
\altaffiltext{3}{Institute of Cosmology \& Gravitation, Dennis Sciama Building, University of Portsmouth, Portsmouth, PO1 3FX, UK.}
\altaffiltext{4}{Departamento de F\'isica Matem\'atica, Instituto de F\'isica, Universidade de S\~ao Paulo, CP 66318, CEP 05314-970, S\~ao Paulo, SP, Brazil}
\altaffiltext{5}{Laborat\'orio Interinstitucional de e-Astronomia, 77 Rua Gal.~Jos\'e Cristino, Rio de Janeiro, RJ - 20921-400, Brazil.}
\altaffiltext{6}{Observat\'orio Nacional, 77 Rua Gal. Jos\'e Cristino, Rio de Janeiro, RJ - 20921-400, Brazil.}
\altaffiltext{7}{Universidade Federal do ABC, Centro de Ci\^encias Naturais e Humanas, Av. dos Estados, 5001, Santo Andr\'e, SP, Brazil, 09210-580}
\altaffiltext{8}{ Instituto de F\'isica Te\'orica, Universidade Estadual Paulista, Rua Dr. Bento T. Ferraz, 271, S\~ao Paulo, SP, 01140-070, Brazil}
\altaffiltext{9}{Department of Physics and Astronomy, University of Utah, Salt Lake City, UT 84112, USA.}
\altaffiltext{10}{Instituto de F\'{\i}sica Te\'orica UAM/CSIC, 28049 Madrid, Spain.}
\altaffiltext{11}{Departamento de F\'{\i}sica Te\'orica, Universidad Aut\'onoma de Madrid, 28049 Madrid, Spain.}
\altaffiltext{12}{Department of Physics and Astronomy and PITT PACC, University of Pittsburgh, Pittsburgh, PA 15260, USA.}
\altaffiltext{13}{Departamento de F\'isica, Facultad de Ciencias, Universidad de La Serena, Cisternas 1200, La Serena, Chile.}
\altaffiltext{14}{Department of Astronomy and Astrophysics, The Pennsylvania State University, University Park, PA 16802.}
\altaffiltext{15}{Institute for Gravitation and the Cosmos, The Pennsylvania State University, University Park, PA 16802.}

\begin{abstract}
We present the first scientific results from the luminous red galaxy sample (LRG) of the extended Baryon Oscillation Spectroscopic Survey (eBOSS) combined with the high-redshift of the previous BOSS sample. We measure the small and intermediate scale clustering from a sample of more than 97,000 galaxies in the redshift range $0.6 < z < 0.9$. We interpret these measurements in the framework of the Halo Occupation Distribution. The bias of this sample of LRGs is $2.30 \pm 0.03$, with a satellite fraction of $13\pm3$\% and a mean halo mass of $2.5\times10^{13}h^{-1}M_{\odot}$. These results are consistent with expectations, demonstrating that these LRGs will be reliable tracers of large scale structure at $z\sim 0.7$. The galaxy bias implies a scatter of luminosity at fixed halo mass, $\sigma_{\log L}$, of 0.19 dex. Using the clustering of massive galaxies from BOSS-CMASS, BOSS-LOWZ, and SDSS, we find that $\sigma_{\log L}=0.19$ is consistent with observations over the full redshift range that these samples cover. The addition of eBOSS to previous surveys allows investigation of the evolution of massive galaxies over the past $\sim 7$ Gyr.
\end{abstract}

\keywords{large scale structure of universe}

\section{Introduction}

Galaxy redshift surveys have been fundamental in advancing our understanding of the universe. The successes of the past decade, varying from 2dFGRS \citep{Cole_2005}, SDSS \citep{Eisenstein_2005, Zehavi_2011}, and BOSS \citep{Anderson_2012}, have spawned even larger investments in mapping the universe through the three-dimensional distributions of galaxies. In this paper, we present the first measurements of the clustering of luminous red galaxies (LRGs) from the extended Baryon Oscillation Spectroscopic Survey (eBOSS; \citealt{eBOSS_Dawson}), the successor program to BOSS (\citealt{Dawson_BOSS}). The eBOSS LRG program has the power to provide reliable measurements of galaxy clustering.

We focus on LRG clustering at small scales ($r \lesssim 20$ $h^{-1}$ Mpc), scales which provide information on the bias of the galaxy sample and how these galaxies are distributed in dark matter halos. The framework in which we interpret the eBOSS data is the Halo Occupation Distribution (HOD). 
This approach describes the bias relation between the galaxies and matter  at the level of ``virialized" dark matter halos which are expected to be in approximate dynamical equilibrium \citep{HOD_Weinberg, Peacock_2000, Seljak_2000, Benson_2000, Martin_2001, Cooray_2002}. In the HOD framework, the key quantity is the probability distribution $P(N|M)$ that a halo of virial mass $M$ contains $N$ galaxies of a given type, along with the relations between the galaxy and dark matter spatial and velocity distributions within halos. Given an HOD and a particular cosmological model, the statistics  of galaxy clustering can be predicted in the sense that the cosmological model determines the properties of the halo distribution, while the HOD specifies how those halos are populated with galaxies.
HOD modeling has been used to interpret clustering in nearly all large-scale galaxy redshift surveys (e.g. \citealt{Zheng_DEEP2, Zheng_2009,  Zehavi_2011, CMASS_Martin, Parejko_LOWZ, Guo_2014}). The HOD results provide physically informative and important information to test theories of galaxy formation and evolution.

One of the key quantities in galaxy formation is the scatter in galaxy luminosity (or stellar mass) at fixed halo mass. Clustering is one of the few methods that is sensitive to the scatter. We will use the HOD to estimate this scatter and compare it to other galaxy samples spanning a redshift range of $z=0.7$ to $z=0.1$. We will show that this scatter is both small (0.19 dex in $\log{L}$) and constant over this redshift range.

Our paper is organized as follows. 
Section 2 briefly describes the eBOSS observations and the definition of our LRG sample. The measurement of clustering is presented in Section 3, along with the comparison with the BOSS result. In Section 4, we interpret our result in the framework of HOD. Finally, the conclusion and discussion of our measurements as well as its implication are given in Section 5. Throughout this paper, the distances are measured in units of $h^{-1}$ Mpc with the Hubble constant $H_{0}$=100 $h$ km s$^{-1}$ Mpc$^{-1}$.
The redshifts are converted to distances by assuming a spatially flat $\Lambda$CDM model with $(\Omega_{m}, h,  \Omega_{b}, \sigma_{8}, n_{s})$ = $(0.29, 0.7, 0.04, 0.8, 0.95)$. The same cosmology is also used for the $N$-body simulations to make mock catalogs. The halos are defined as the spherical overdensity masses which are 200 times the background density.

\section{Observations and data}\label{sec:data}

Motivated by the success of BOSS \citep{Bolton_2012, Dawson_BOSS, Eisenstein_2011, APO_Gunn, Smee_2013}, eBOSS will explore a larger volume and higher redshift of the universe (\citealt{eBOSS_Dawson}). As a six-year program, the primary scientific goals of eBOSS are to provide the first high precision measurements of baryon acoustic oscillations (BAO) and redshift space distortions (RSD) in the redshift range $0.6 < z < 2.0$ \citep{Zhao_eBOSS}. Measurements of the expansion history in this redshift range contain important information about the transition from cosmic deceleration to acceleration. Here, we focus on the eBOSS LRG sample, which extends the BOSS galaxy sample to higher redshift, probing the range $0.6<z<1.0$ with a target density of $~60$ deg$^{-2}$. The LRG target selection is based on $ugriz$ (\citealt{Fukugita_1996}) SDSS imaging data combined with infrared photometry from Wide-Field Infrared Survey Explorer (WISE; \citealt{WISE_Wright}). The use of infrared data allows selections of fainter optical targets at higher redshift while minimizing stellar contamination of the sample. A full description of the target selection algorithm, including tests for systematics, is presented in \citet{LRG_Prakash}.

The eBOSS LRG target selection imposes a bright limit of $i=19.9$, making the eBOSS sample nearly complementary to the BOSS CMASS sample, which used SDSS imaging only to probe the redshift range $0.4<z<0.7$ (see details in \citealt{Reid_2016}). For the clustering analysis in this paper, we combine the eBOSS LRGs with the high-redshift tail of the CMASS sample. The motivation for this combination is two fold: (1) HOD analysis typically assumes that a sample of galaxies is complete, in the sense that it includes all galaxies above some mass or luminosity threshold. (2) The cosmology analysis with eBOSS is likely to merge the two catalogs; this increases the density of the sample without decreasing the median redshift. Like other LRG selections, our combined sample of eBOSS+BOSS galaxies is not a complete sample, either in terms of luminosity or stellar mass. Color cuts will introduce some incompleteness, while the flux limit of the target selection will create incompleteness at the higher redshift region of the sample. The completeness of BOSS LRG samples has been quantified by \cite{leauthaud_etal:16} and \cite{tinker_etal:16_boss} using ancillary datasets to augment the BOSS samples. Such samples do not currently exist for eBOSS, thus the analysis presented here comes with the caveat that the measured clustering and derived halo occupation may be biased relative to a stellar-mass complete sample. We will discuss this in the context of HOD modeling in  section 4.

\begin{figure}[htbp]
\begin{center}
\includegraphics[width=9cm, height=6.5cm]{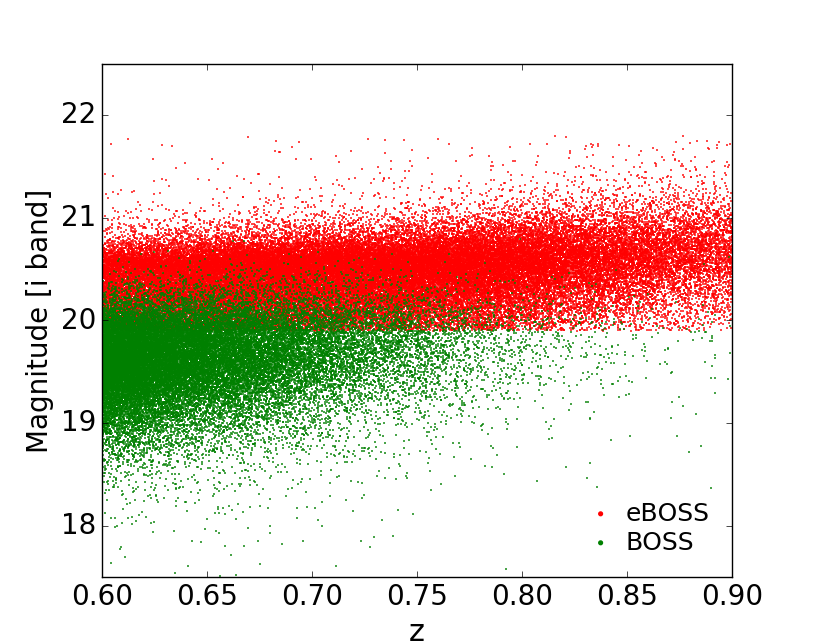}
\caption{The distribution of the $i$-band model magnitude after the correction of galactic extinction $versus$ redshift for the eBOSS (red) and BOSS (green) samples.}
\label{Mag_z}
\end{center}
\end{figure}

Figure \ref{Mag_z} shows the distribution of $i$-band magnitude as a function of redshift for both BOSS-CMASS and eBOSS LRGs. At $z<0.75$, the complementarity of the eBOSS and BOSS samples is clear, with BOSS populating the bright end of the distribution. The slight overlap between BOSS and eBOSS galaxies is due to the bright limit in eBOSS using ``model" magnitudes, while the faint limit in CMASS was enforced with ``cmodel" magnitudes (see \citealt{Stoughton_2002, Abazajian_2004}, for details of the magnitudes and further discussion). The smaller scatter between these two quantities causes some overlap in the $i$-band distribution. At $z>0.75$, the combined sample is dominated by eBOSS galaxies due to the flux limit of BOSS.

\begin{figure}[htbp]
\begin{center}
\includegraphics[width=9cm, height=6.5cm]{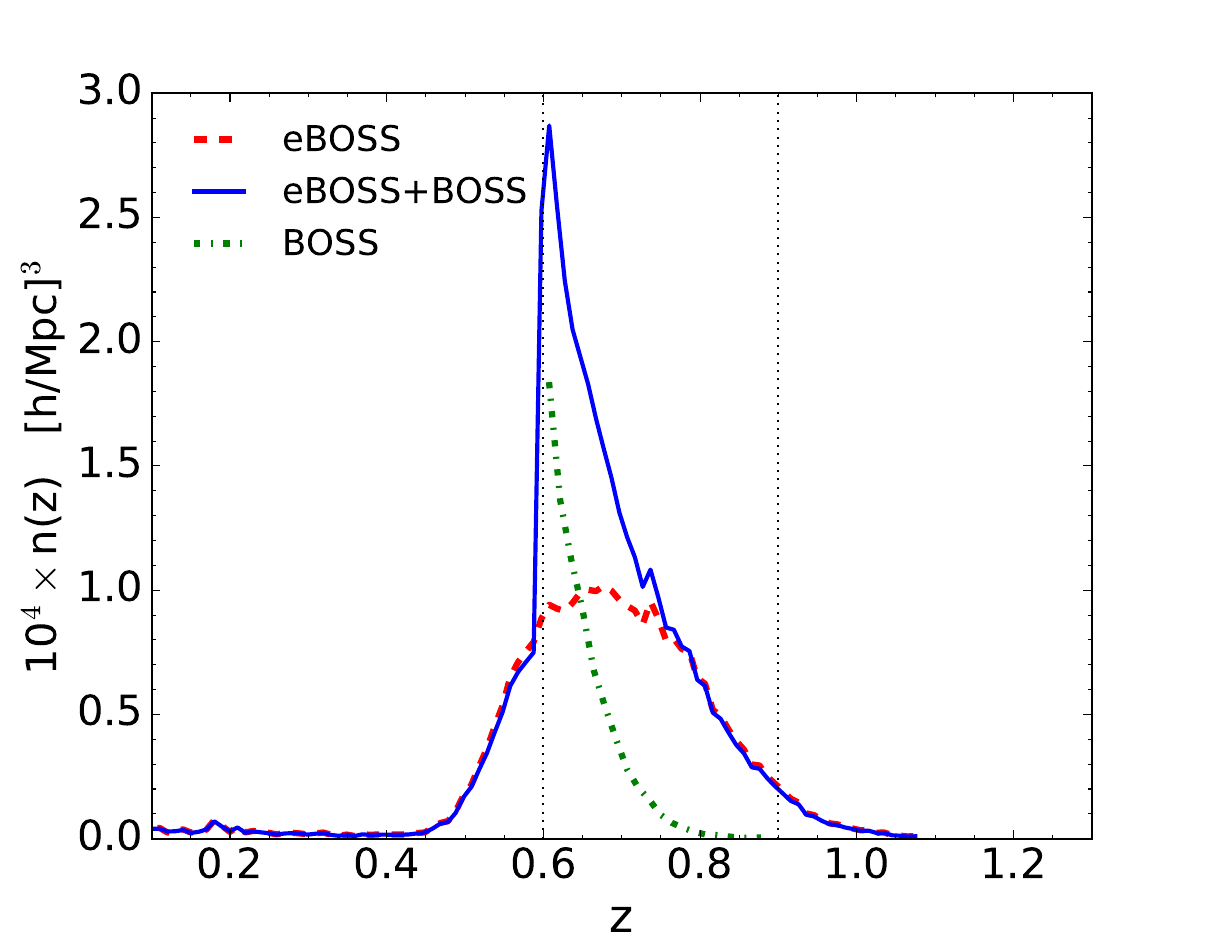}
\includegraphics[width=9cm, height=6.5cm]{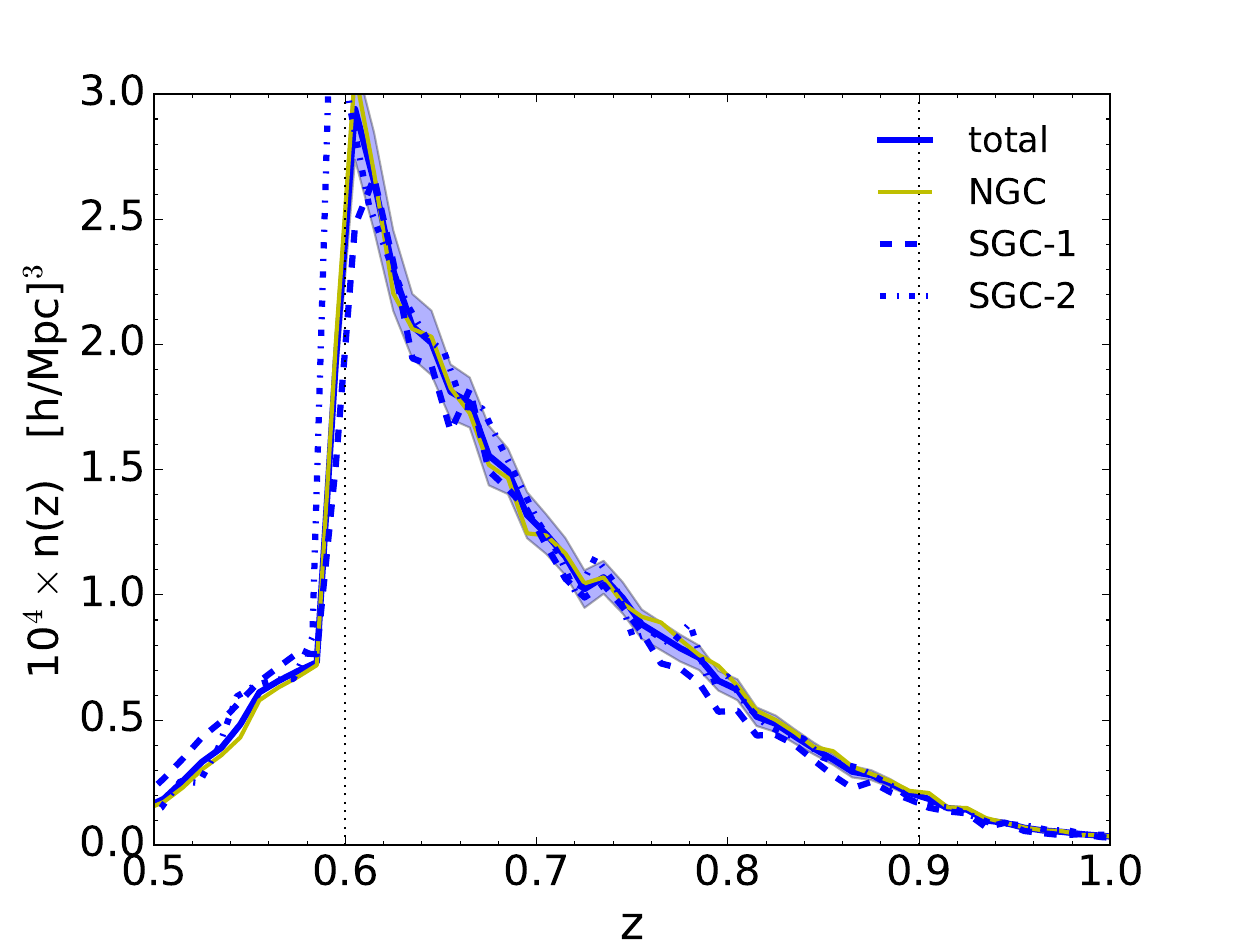}
\caption{$Top$ $panel: $ The number density of the galaxies for the sample described in the text: eBOSS LRGs (red dashed), BOSS (green dot-dashed) and eBOSS+BOSS (blue solid). The data used for the clustering measurement are restricted between the vertical dashed lines at $z=0.6$ and $z=0.9$. $Bottom$ $panel:$ The number density of the galaxies in the eBOSS+BOSS LRG sample, the three patches in NGC and SGC are plotted separately, the shaded region in the restricted redshift range is the $2\sigma$ error estimated from the mock catalogs. The result shows that the data in different areas are consistent with each other.}
\label{n_z}
\end{center}
\end{figure}

The space density of the eBOSS, BOSS, and the combined sample is shown in the the top panel of Figure \ref{n_z}. Dotted lines indicate our fiducial redshift range. Within this range, the fraction of all galaxies that are eBOSS LRGs is about $60\%$. The combination of the eBOSS sample with the flux-limited tail of the CMASS distribution makes for a highly asymmetric $n(z)$, but we will demonstrate in Section \ref{sec:HOD} that our halo occupation results are insensitive to the exact details of the galaxy number density. 
In the bottom panel of Figure 2, we also plot the galaxy number densities of the eBOSS+BOSS LRG sample in different spatial areas, as shown in Figure 3. The consistency between these different patches shows that these data are compatible with each other. As we demonstrate below, the clustering measurements are also consistent between hemispheres, implying that a joint analysis is sufficient for a HOD investigation.

\begin{figure}[htbp]
\begin{center}
\includegraphics[width=9cm, height=6cm]{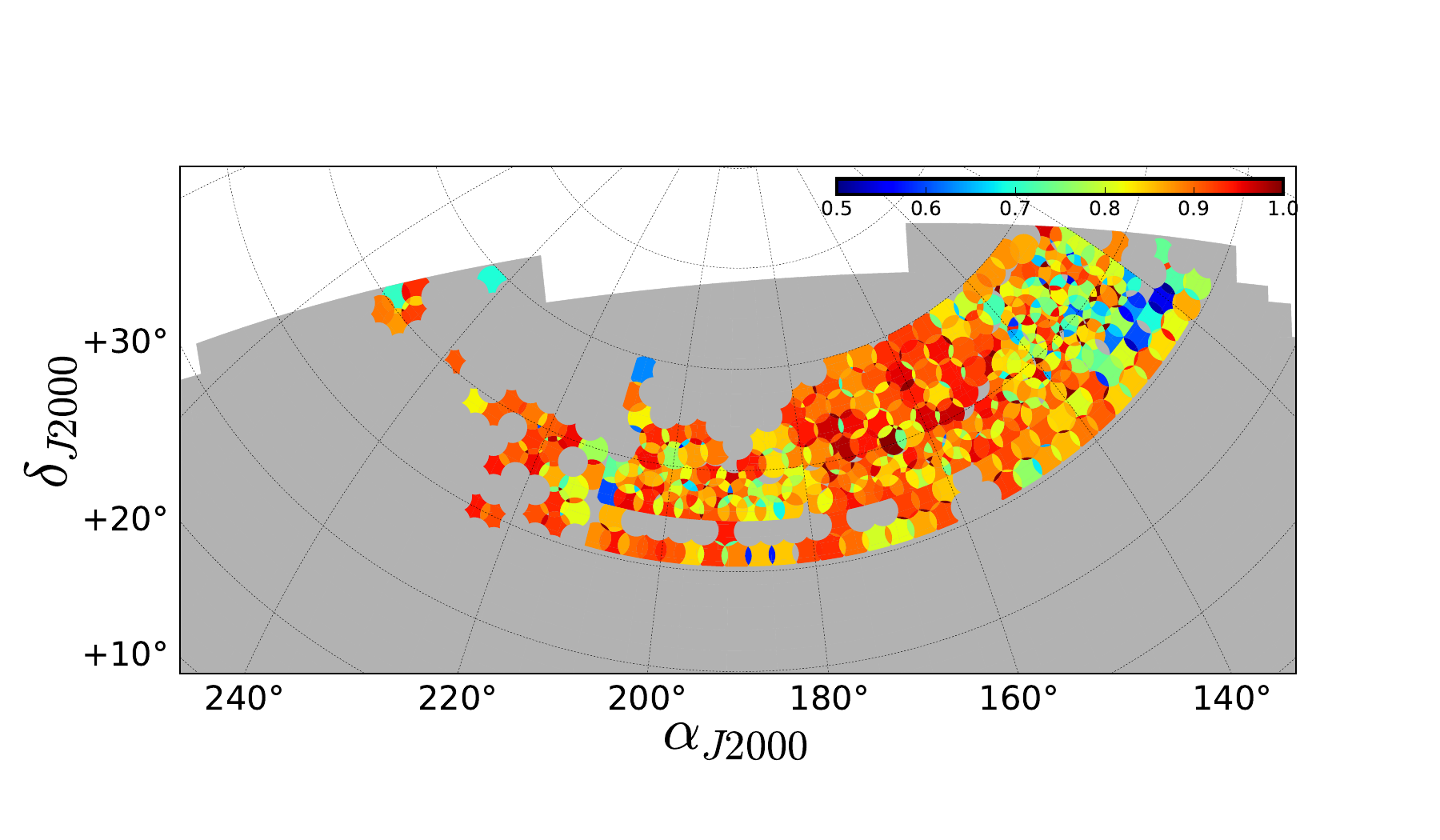} \\
\includegraphics[width=9cm, height=5cm]{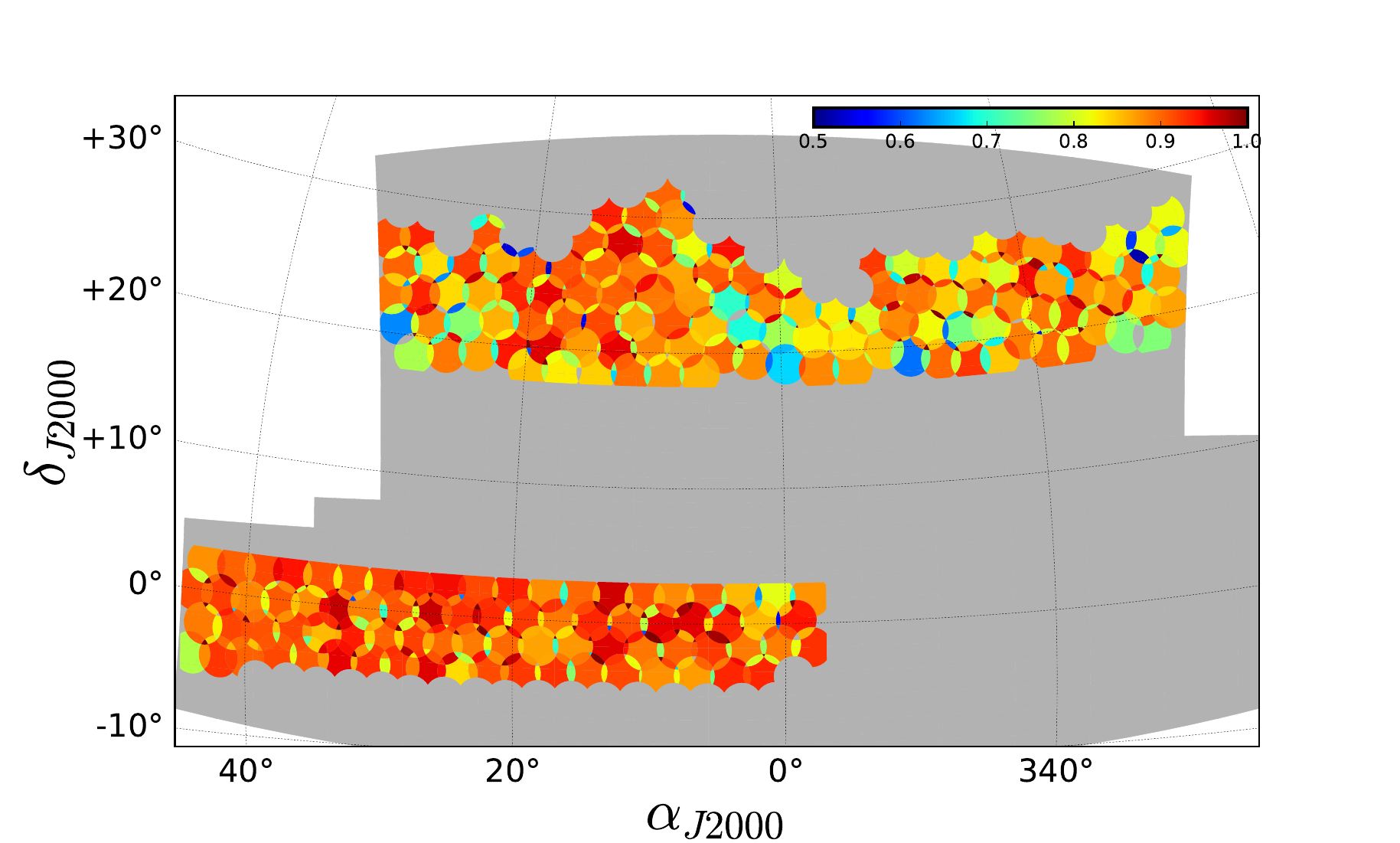}
\caption{The sky coverage of the galaxy sample used in this analysis, in the Lambert azimuthal equal-area projection. The light grey region shows the expected total footprint of the survey, while the colors indicate the completeness in each sector. The mean completeness in each sector is 0.86, the weighted area of the current footprint is 888 deg$^2$ for NGC (top) and 807 deg$^2$ for SGC (bottom) respectively, the two regions in SGC are separated as SGC-1 ($\delta_{J2000}>10^{\circ}$) and SGC-2 ($\delta_{J2000}<10^{\circ}$).}
\label{sky}
\end{center}
\end{figure}

The clustering measurements in this paper are based on the eBOSS DR14 LRG data taken prior to May 2016. This sample yields a total number of spectra of 110,000 and an areal coverage of 1591 deg$^{-2}$. We restrict the data to the sectors with a completeness greater than 0.5 and then select the galaxies in our redshift range. The overall completeness in each sector is defined as 
\begin{equation}
C = \frac{N_{\text{spec}}+N_{\text{cp}}+N_{\text{BOSS}}+N_{\text{BOSS}_{\text{cp}}}}{N_{\text{targ}}-N_{\text{star}}-N_{\text{knocks}}+N_{\text{BOSS}}+N_{\text{BOSS}_{\text{cp}}}},
\end{equation}
where $N$ is the number of objects in the sector, $\mathsf{spec}$ denotes galaxies with good eBOSS spectra, $\mathsf{cp}$ denotes objects with no spectra, because they were too close to another LRG target to assign a fiber---the well-known ``fiber collision" effect, $\mathsf{BOSS}$ denotes BOSS galaxies with spectra, $\mathsf{BOSS}_{\text{cp}}$ has the same meaning as $\mathsf{cp}$ but in the BOSS-CMASS sample, $\mathsf{targ}$ denotes targets, $\mathsf{star}$ denotes spectroscopically confirmed stars, and $\mathsf{knocks}$ denotes knockouts from higher priority targets which we will discuss in more detail presently. In the analysis, we define ``good" eBOSS spectra as follows,
\begin{eqnarray}
&&(\text{1}) \text{SPECPRIMARY} == 1, \quad \mathsf{AND} \nonumber \\
\Big( && (\text{2a}) \text{ZWARNING}\_\text{NOQSO} ==0, \quad \mathsf{OR}   \\ \nonumber
&&(\text{2b}) \text{ZWARNING}\_\text{NOQSO}==2^2 \mathsf{AND}  \nonumber \\
& & 0.005 < \text{RCHI2DIFF}\_\text{NOQSO}  < 0.01 \Big) \nonumber \\  \nonumber
\end{eqnarray}
These parameters are the flags in the eBOSS catalogue:  $\text{SPECPRIMARY}$ identifies the best spectrum among multiple observations, $\text{ZWARNING}\_\text{NOQSO}$ lists potential problems with the redshift fit and a value of 0 denotes no obvious problems, $\text{RCHI2DIFF}\_\text{NOQSO}$ is the difference of reduced $\chi^2$ between the best-fit and second best-fit templates. 
The third condition is used to relax the threshold of $\text{ZWARNING}\_\text{NOQSO}=2^2$, since in \cite{eBOSS_Dawson} it was shown that the catastrophic failure rate is still below 1\% for $\text{RCHI2DIFF}\_\text{NOQSO} > 0.005$. 
Future analysis of clustering with the eBOSS LRG sample will likely use the redshift estimates derived from a new spectroscopic classification algorithm (\citealt{Hutchinson_16_redmonster}).  The new routine is based on a least squares fit against discrete, physically-motivated spectral templates rather than against a linear combination of templates derived from principal component analysis as was done in BOSS redshift classification (\citealt{Bolton_2012}).  The new redshift classification algorithm has been shown to produce a higher fraction of reliable redshift estimates, particularly in the presence of stellar contamination and low signal-to-noise spectra.
The above approach increases the redshift success rate by about 15\%. These criteria yield an eBOSS LRG sample of 62,000 galaxies, and the redshift success rate is 84\%. The stellar fraction in the spectroscopic sample is found to be about 11\%. We restrict the CMASS galaxies to the same footprint as the eBOSS sample resulting in an eBOSS+BOSS sample of 97,000 galaxies. We summarize the basic statistics of the eBOSS+BOSS sample in Table I, including the galaxy numbers, the space density, completeness and stellar contamination.
Figure \ref{sky} displays the sky coverage of the eBOSS LRG sample color-coded by completeness. The North Galactic Cap (NGC) and South Galactic South (SGC) are analyzed jointly for simplicity. 
The area covered by the survey and the angular completeness of each sector is tracked by the $\tt{MANGLE}$ software \citep{Swanson_Mangle}. 

\begin{table}
\caption{The statistics of the eBOSS+BOSS LRG sample}
\begin{center}
\begin{tabular}{rcccc}
\\  & Total & NGC & SGC-1 & SGC-2 \\ 
 \cline{1-5}
 \\
$N_{\rm{gal, total}}$  &  97073  &   51388  &  25649  & 20036 \\
$N_{\rm{gal, BOSS}}$  & 34924 &  18637  & 8625  & 7662  \\
$N_{\rm{gal, eBOSS}}$  &   62149 &  32751  & 17024  & 12374 \\
area (deg$^2$)  & 1695.5 &  888.5 & 482.1 &  324.9 \\
stellar comtamination  & 6.9\% & 4.6\% & 12.2\% \footnote{The stellar contamination is higher due to this region being closest to the galactic plane. But the clustering is not affected.}  & 4.3\% \\
completeness & 0.863 & 0.862 & 0.848 & 0.888  \\
\\
\cline{1-5}
\end{tabular}
\end{center}
\label{tab:stats}
\end{table}

We apply additional masks to the data to account for various systematics. During fiber assignment, LRGs are only given access to fibers after all other targets have been through fiber allocation. Thus, there is a significant amount of area that is ``not viewable" from the point of view of the LRGs due to fiber collisions --- the limit that two fibers cannot be closer than 62$''$ on a given plate; see \citet{eBOSS_Dawson} for full details. LRG targets that are within the collision radius of a high priority target are designated knockouts (the collision of a LRG with another LRG will be discussed later). Some knockouts are recovered in plate overlaps, but in total roughly 10\% of the LRG footprint is eliminated due to this effect. We create a collision priority mask to remove both targets and randoms from this area.
Bright stars in WISE can also impact target selection. In our fiducial results, we do not use the bright star mask, but we demonstrate that it has negligible effect on our clustering measurement in Appendix \ref{bright}.

\section{Clustering measurements}\label{sec:clustering}

The primary tool to study the statistics of the galaxy distribution is the two-point correlation function $\xi(r)$, which measures the excess probability of finding galaxy pairs over a random sample of points as a function of separation $r$ \citep{Peebles1980large}.
In order to account for the redshift space distortions caused by the galaxy peculiar velocities, it is convenient to calculate the correlation function on a two-dimensional grid of pair separations perpendicular ($r_{p}$) and parallel ($\pi$) to the line of sight. 
For a pair of galaxies with redshift space positions $\textbf{s}_{1}$ and $\textbf{s}_{2}$, the dependence of the correlation function is only through $\textbf{s}=\textbf{s}_{1}-\textbf{s}_{2}$ and the orientation of $\textbf{s}$ relative to the line-of-sight. In this case, we may write the $\xi(r)$ as $\xi(\pi, r_{p})$ through the relation
\begin{equation}
\pi = \frac{\textbf{s}\cdot \textbf{l}}{|\textbf{l}|}, \quad r_{p}^2=\textbf{s}\cdot \textbf{s}-\pi^2, \quad r^2 = \textbf{s}\cdot \textbf{s},
\end{equation}
with $\textbf{l}=(\textbf{s}_{1}+\textbf{s}_{2})/2$ (e.g. \citealt{Davis_1983, Fisher_1994}).

The calculation of the correlation function from the galaxy sample is through the estimator (\citealt{LS_1993})
\begin{equation}\label{eq:LS}
\xi(r_{p}, \pi)=\frac{DD-2DR+RR}{RR},
\end{equation}
where $DD$, $DR$, and $RR$ are suitably normalized numbers of (weighted) data\textendash data, data\textendash random, and random\textendash random pairs in each separation bin. Note that a FKP type radial weighting is not applied here, as it has no impact on the clustering at this scale (\citealt{CMASS_Martin, Parejko_LOWZ}).
We generate the random catalogs in the survey area which satisfies the completeness threshold, and assign a weight of 1 to all of these randoms. The redshifts of these randoms are selected randomly from redshifts in the data sample. We subsample the randoms in each sector to match the incompleteness of the spectroscopic sample.

In order to mitigate the effect of redshift space distortion and examine the real space correlation function, we compute the projected correlation function from $\xi(r_{p}, \pi)$
\citep{Davis_1983}
\begin{equation}
w_{p}(r_{p})=2\int^{\infty}_{0}d\pi \xi(r_{p}, \pi).
\end{equation}
In practice, the integral of $\pi$ can be up to 80 $h^{-1}$Mpc, which is large enough to include most of the correlated pairs and produce a stable result. The measurement of $w_{p}(r_{p})$ is achieved with 10 equally spaced bins in $\log{r_{p}}$ from 0.2 $h^{-1}$Mpc to 60 $h^{-1}$Mpc. 

Fiber collisions between LRG-LRG pairs reduce the spectroscopic completeness by $\sim5\%$\footnote{This is a distinct effect from knockouts, where LRGs cannot be assigned fibers due to collisions with other---uncorrelated---samples of targets. This effect is specifically caused by the collision between two LRGs.}, and these collisions have an impact on both the measured large-scale bias and the small scale clustering. 
We correct this effect by combining two different weights: (1) upweighting galaxies which have a fiber assigned in the collided-pairs and (2) reconstructing the correct galaxy pair counts in scales smaller than 62$''$. The first weighting scheme is similar to the ``nearest-neighbor method" and corrects for the impact of collisions on the bias\citep{Zehavi_nearest, Zehavi_2005}. The second scheme corrects the clustering amplitude at small scales by using the ratio of angular correlation functions \citep{Hawkins_2003}
\begin{equation}\label{eq:F}
F(\theta)=\frac{1+w_{z}(\theta)}{1+w_{t}(\theta)},
\end{equation}
where $w_{z}(\theta)$ is the angular correlation function of galaxies drawn from the ``spectroscopic" sample which has fibers assigned, and $w_{t}(\theta)$ is the angular correlation function for the entire photometric sample. 

\begin{figure}[htbp]
\begin{center}
\includegraphics[width=9cm, height=6.5cm]{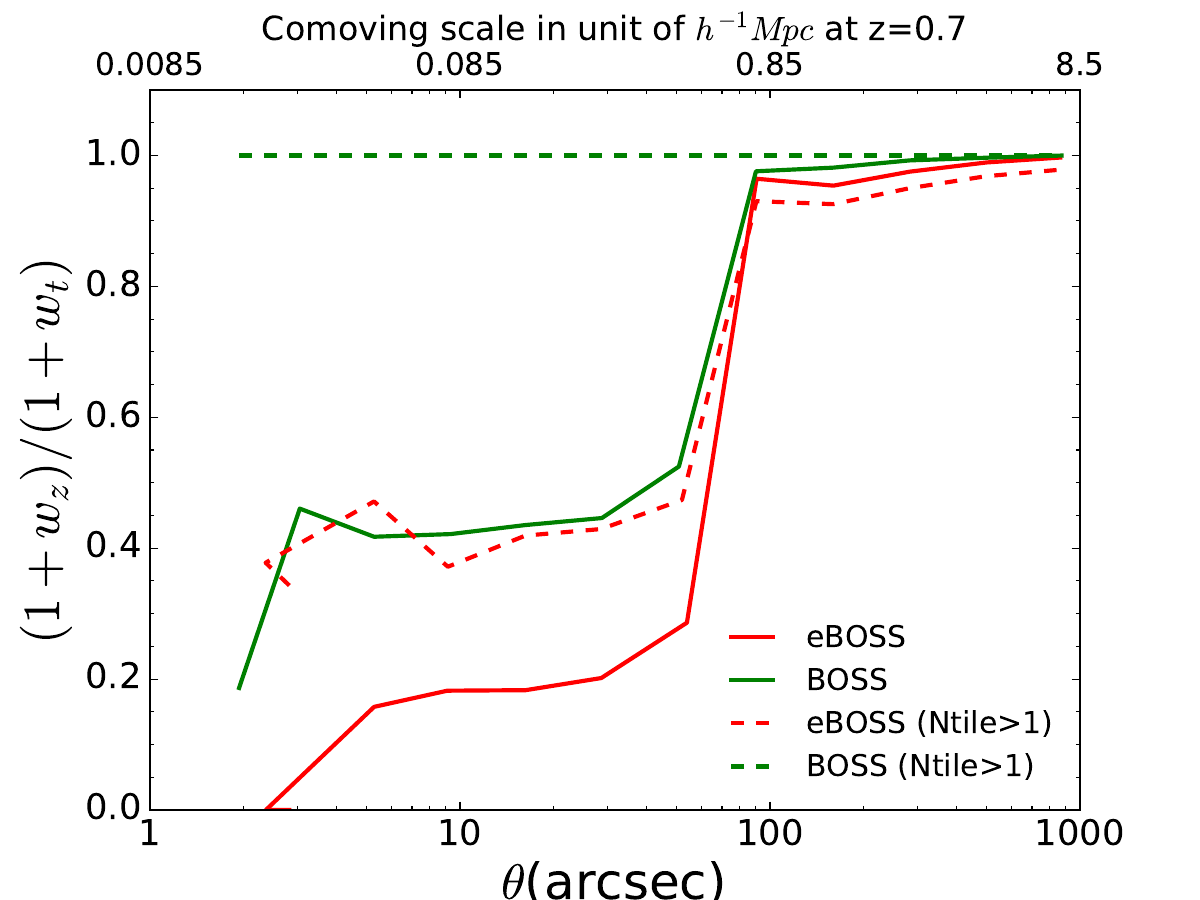} \\
\caption{The angular correction (Eq.\ref{eq:F}) for eBOSS (red) and BOSS (green) samples in the pair counts to calculate the correlation function. $N_{\rm{tile}}>1$ refers to sectors that are observed more than once. This quantity is used to weight the galaxy pairs to account for the loss of pairs due to collisions.}
\label{theta}
\end{center}
\end{figure}

The quantity $1+w(\theta)$ is proportional to the number of pairs at angle $\theta$, thus we weight each DD pair in Eq. (\ref{eq:LS}) by $1/F(\theta)$ to account for the loss of pairs due to collisions.
Figure \ref{theta} presents this angular correction for both eBOSS and BOSS galaxy samples used in our analysis. The ratio is close to unity above the fiber collision scale but depressed significantly at separations below this scale. 
To interpret these data we start with results from BOSS. For fiber allocation in BOSS, the mandate was to place a fiber on every galaxy possible--- i.e., to achieve 100\% completeness in the `decollided' set \footnote{The decollided set contains all targets that are not within collision groups (groups of targets that lie within 62$''$ of one another), combined with the subset of collided targets that can be assigned fibers on a single plate (\citealt{eBOSS_Dawson}).}. 
Thus, in areas of the survey covered by more than one tile, all collisions were resolved by observing one galaxy on each plate. Because 40\% of BOSS was covered by more than one tile, the value of $F$ at $\theta < 62''$ is $~0.4$. The value of $F$ for eBOSS galaxies is substantially smaller below the collision scale, in spite of the fact that the multi-tile coverage is nearly the same. Indeed, when measuring $F(\theta)$ in regions of eBOSS covered by more than one tile, $F(\theta)$ is still substantially below unity.

The reason for the different results between BOSS and eBOSS lies in the fiber allocation priorities. In BOSS, the goal of 100\% completeness in the decollided set was met at the expense of some unused fibers, which totalled $~7\%$. To maximize fiber usage in eBOSS, the goal of 100\% decollided completeness was relaxed for the LRGs (but only for the LRGs). Due to fluctuations in the density of higher priority targets, the number of LRG fibers varied from plate to plate. Thus, in some plates, there exist more LRG targets than available fibers. This effects a small fraction of area; $~90\%$ of the eBOSS footprint placed fibers on $\geq 90\%$ of available LRG targets (cf. Figure 4 in \citealt{eBOSS_Dawson}). However, the fiber allocation algorithm prioritizes galaxies in the decollided set. Thus, if a plate runs out of fibers before all available LRGs could be assigned, the set of LRGs left unassigned are preferentially in pairs. Correcting for this effect, fortunately, is identical to our standard method of correcting for fiber collisions; this result is shown explicitly on mock data in Appendix \ref{mock}.

\begin{figure}[htbp]
\begin{center}
\includegraphics[width=9cm, height=7.5cm]{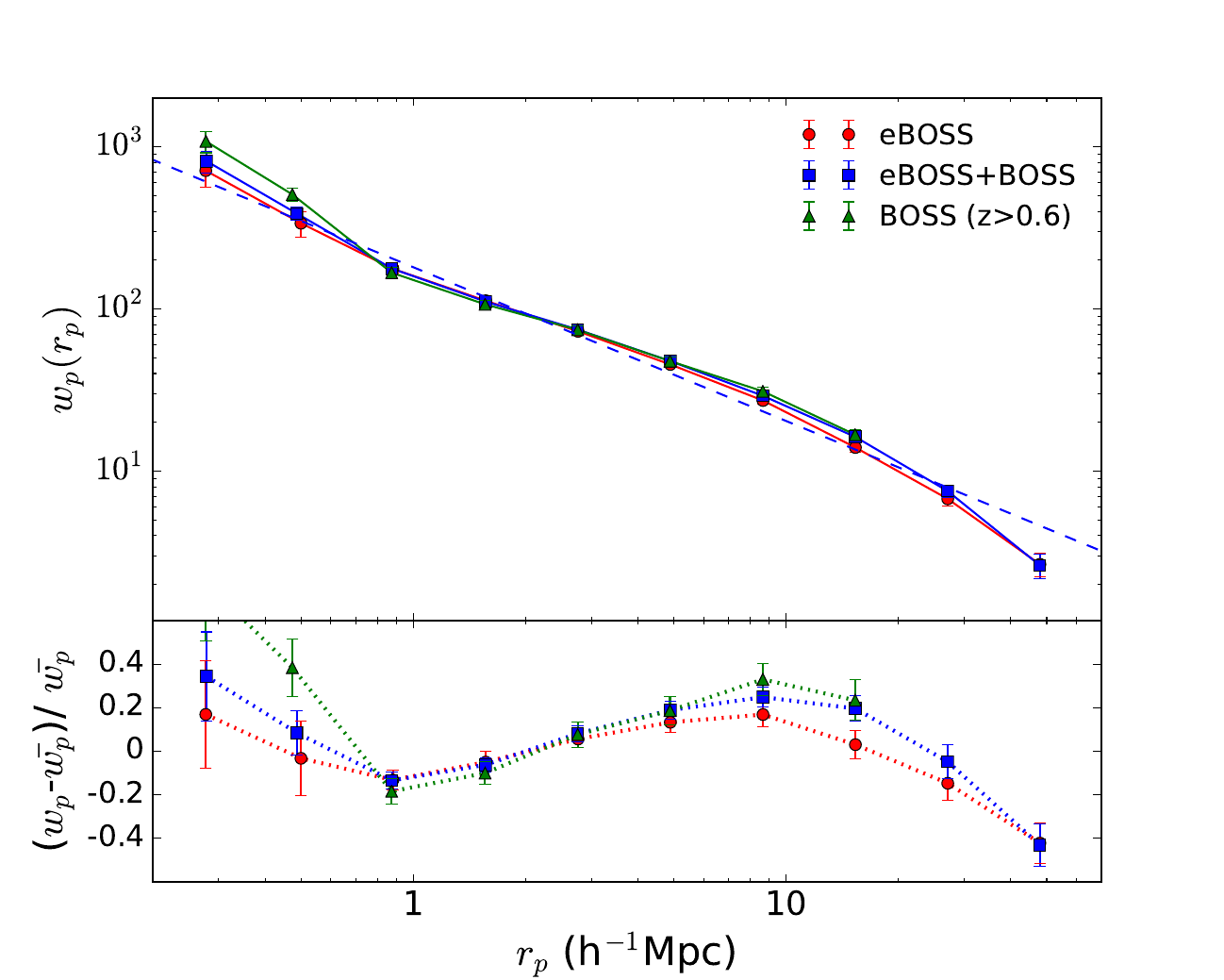}
\caption{$Top$ $panel$: Projected correlation function for the eBOSS, BOSS and eBOSS$+$BOSS LRG samples. The dashed line corresponds to the best-fit power law for eBOSS LRGs $w_{p}(r_{p}) \propto r_{p}^{1-\gamma}$ with $\gamma\sim 1.95$. $Bottom$ $panel$: The fractional difference for the two samples with respect to the best-fit power law function. Note that the two data points at the largest scale for BOSS $w_{p}$ are not shown, because they are negative due to the sample variance which is mainly introduced by restricting the BOSS CMASS galaxies within the eBOSS footprint.}
\label{wp_obs}
\end{center}
\end{figure}

\begin{figure}[htbp]
\begin{center}
\includegraphics[width=9cm, height=8.5cm]{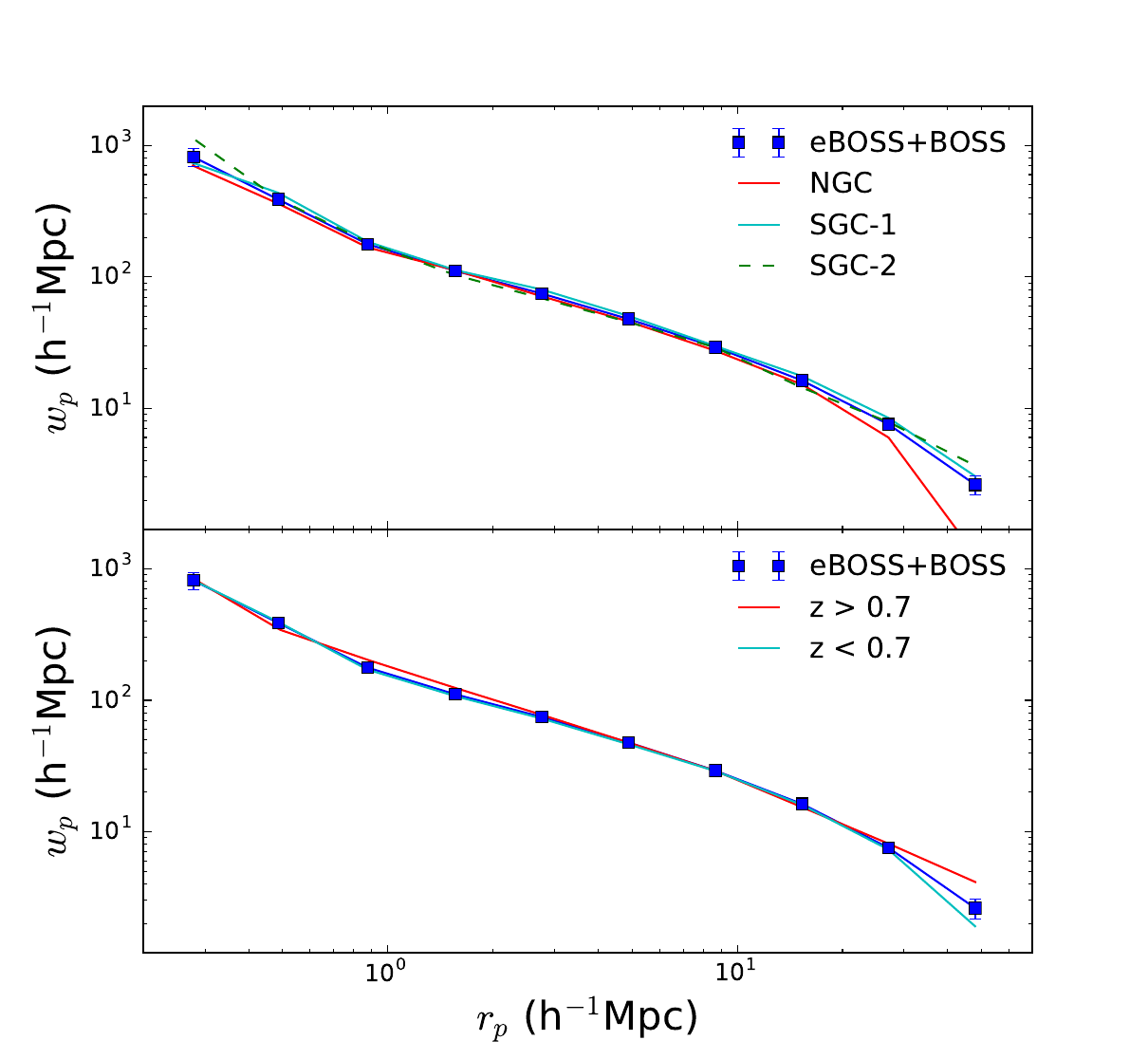}
\caption{$Top$ $panel$: Projected correlation function for the regions NGC, SGC-1 and SGC-2 of the eBOSS$+$BOSS LRG sample, together with the overall result. The consistency between these subsamples are measured by $\chi^{2}=(w_{p1}-w_{p2})C_{tot}^{-1}(w_{p1}-w_{p2})$, determined by taking the difference between these subsamples, where the covariance is determined by scaling with the number of galaxies in the subsample. The result is 6.8 for NGC and SGC1, 17.0 for NGC and SGC2, 9.9 for SGC1 and SGC2, with 10 data points. $Bottom$ $panel$: The projected correlation function of the high- and low-$z$ samples, the corresponding $\chi^{2}=17.1$. The results show that the LRG sample in different areas and redshift bins are compatible with each other, and therefore motivate us to analyze the data in a joint way.}
\label{wp_split}
\end{center}
\end{figure}

Our measurement of $w_{p}(r_{p})$ is shown in Figure \ref{wp_obs} for CMASS, eBOSS LRGs, and the combined sample. 
Note again that we restrict the CMASS sample to be within the same eBOSS survey area and redshift range. The angular completeness and the radial selection function are calculated for eBOSS and eBOSS+BOSS independently, and the angular upweighting correction is also applied to eBOSS and BOSS LRGs separately. 
Specifically, in the combined sample, a BOSS-BOSS pair at $\theta<62''$ is upweighted by 2.64, while an eBOSS-eBOSS pair is upweighted by 5.1, and all eBOSS-BOSS cross pairs are not upweighted because there are no collisions between surveys.
The clustering measurements from eBOSS are in agreement with earlier measurements of massive galaxies at lower redshift \citep{CMASS_Martin, Parejko_LOWZ}. 
In the top panel of Figure 6, we present the clustering measurements of the eBOSS+BOSS LRG sample in different regions. These results show consistency with the combined sample. We also subdivide the redshift range into a low-$z$ and a high-$z$ half at about $z=0.7$ such that the two subsamples have nearly equal number of galaxies. The result shown in the bottom panel of Figure 6 reveals no significant difference between the two samples, and thus motivates our analysis of these data as a single sample.

\begin{figure}[htbp]
\begin{center}
\includegraphics[width=9cm, height=7cm]{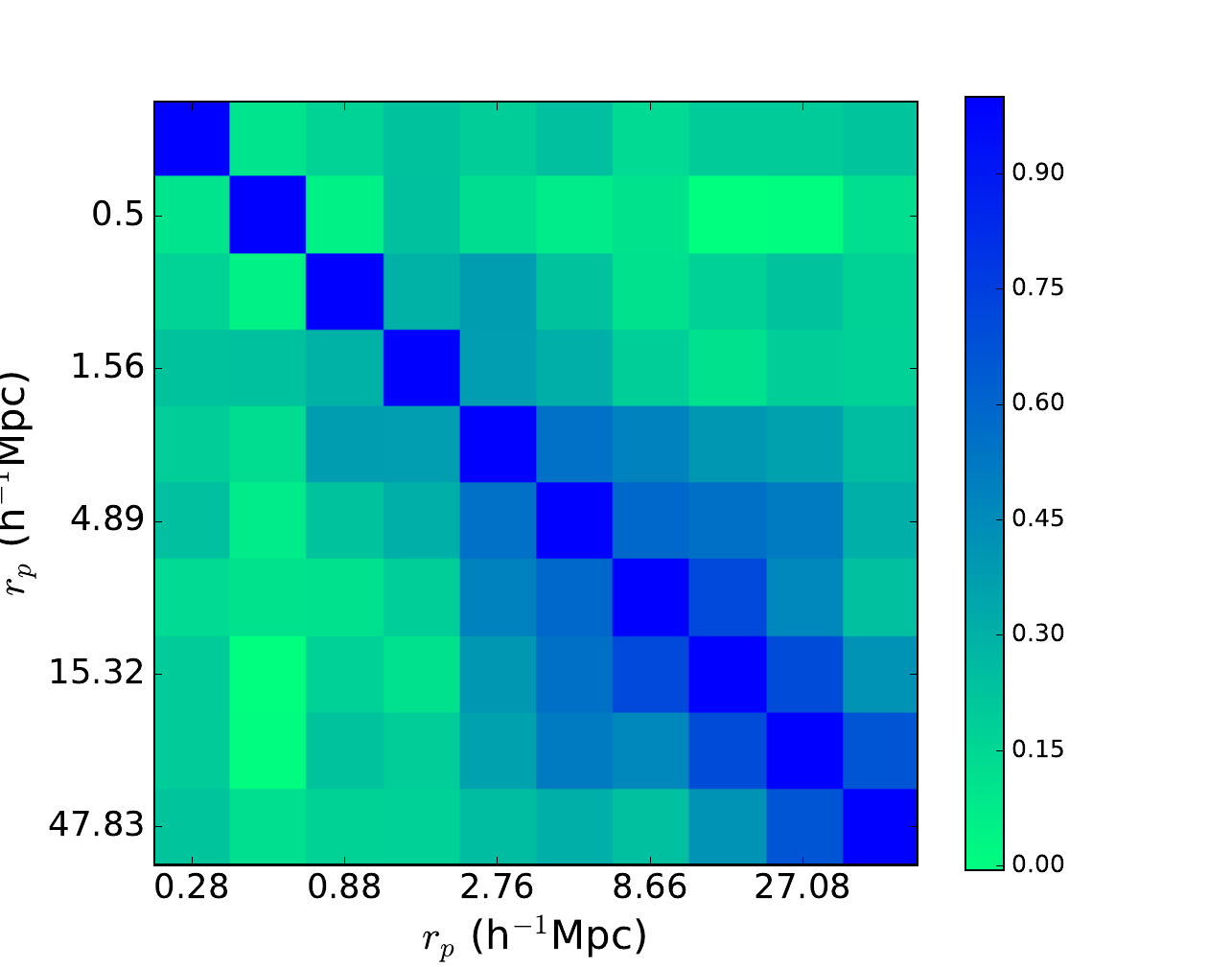}
\caption{The correlation matrix for the clustering measurements from the eBOSS+BOSS LRG sample. It is calculated from 100 independent mock catalogs by the use of the particle mesh method. This matrix is used to perform the HOD analysis in Section \ref{sec:HOD}}
\label{cov}
\end{center}
\end{figure}

The errors in the clustering measurements can be estimated in multiple ways \citep{Norberg_2009}. The eBOSS survey is far from complete, therefore the relatively small sky coverage and the irregular geometry (Figure \ref{sky}) may introduce some difficulties in calculating the covariance matrix from the resampling methods, such as jackknife and bootstrap \citep{CMASS_Martin}. We compute the covariance from 100 independent mock catalogs created from the quick particle mesh method (QPM; \citealt{QPM_White}). 
These mock catalogs have the same angular selection function and $n(z)$ as the data. They do not include fiber collisions or fiber allocation effects, so we increase the variance from the mocks by $1/F(\theta)$ at $\theta<62''$ to account for the larger shot noise in the data at small scales. In practice, we require an HOD model to make mock catalogs: we perform a `first-pass' HOD analysis on the data (see \S \ref{sec:data}) assuming constant fractional errors in $w_{p}(r_{p})$. The resulting HOD is used to populate the mock catalogs, the output of this process is a mock with constant number density of galaxies, we then subsample them to match the observed $n(z)$, which are then used to perform our final HOD analysis on the data. This procedure is advantageous for a number of reasons: it is simple; this makes the clustering constant across the redshift range, meeting observations; the sample variance and shot noise on the $n(z)$ between different mock realizations can be properly modeled\footnote{Specifically, we subsample the mock galaxies such that the {\it mean} $n(z)$ of all mocks matches the observed $n(z)$, thus mock-to-mock variations in $n(z)$ are preserved.}. This method has been used in the LRG clustering analysis, e.g. \citealt{CMASS_Martin, Parejko_LOWZ}.
The resulting correlation matrix for eBOSS+BOSS sample is presented in Figure \ref{cov}. The error bars of $w_{p}(r_{p})$ measurements in Figure \ref{wp_obs} are related to the diagonal elements of the covariance matrix.
As expected, the $w_{p}(r_{p})$ data are highly correlated at $r_{p}\gtrsim2$ $h^{-1}$Mpc, where pairs of galaxies come from two distinct halos, while at smaller scales galaxy pairs reside in a single halo which is dominated by uncorrelated shot noise.

\section{Analysis}
\subsection{HOD modeling}\label{sec:HOD}

We interpret the observed clustering of galaxies in the framework of the HOD which approaches the problem of galaxy bias statistically. In its most basic form, the HOD constructs a probability distribution $P(N|M)$: the probability that a halo of mass $M$ contains $N$ galaxies of a given class. Here, the class of galaxies is the combined eBOSS+BOSS sample. Because the clustering, abundance, and interior structure of dark matter halos is well known from simulations, specifying $P(N|M)$ essentially provides a complete description of the spatial distribution of galaxies. 
For HOD parameterization, it is customary to separate the contribution of the central galaxies from that of the satellite galaxies with the mean occupancy of halos:
\begin{equation}\label{N_tot}
N(M)=\langle N_{\text{gal}}(M) \rangle = N_{\text{cen}}(M)+N_{\text{sat}}(M).
\end{equation}
The mean number of the central galaxies in each halo is modeled with a smooth transition between 0 and 1 galaxy:
\begin{equation}\label{N_cen}
N_{\text{cen}}(M) = \frac{1}{2} \left[1+\text{erf} \left(\frac{\log{M}-\log{M_{\text{min}}}}{\sigma_{\log{M}}}\right)\right],
\end{equation}
and the mean number of satellite galaxies is parameterized as
\begin{equation}\label{N_sat}
N_{\text{sat}}(M) = \left(\frac{M}{M_{\text{sat}}}\right)^{\alpha}\exp{\left(-\frac{M_{\text{cut}}}{M}\right)} N_{\text{cen}}(M).
\end{equation}
Multiplying the central galaxy occupation function in this form guarantees that the satellite occupation terminates at a mass higher than the central occupation cutoff. In this HOD model, $M_{\text{min}}$, $\sigma_{\log{M}}$, $\alpha$, $M_{\text{sat}}$ and $M_{\text{cut}}$ are the free parameters to be fit by observations which include both $w_{p}(r_{p})$ and the observed number density of galaxies. 
Briefly, $M_{\text{min}}$ is the mass at which half the halos have a central galaxy,
$\sigma_{\log{M}}$ physically relates to the scatter of halo mass at fixed galaxy luminosity, 
$\alpha$ is the power-law index for the mass dependence of the number of satellites, $M_{\text{sat}}$ is a typical mass for halos to host one satellite, and $M_{\text{cut}}$ allows for the cutoff in the satellite occupation function to vary with halo mass.
Different functional forms of the HOD parameterization have been applied in the literature, but the model in Eqs.\ref{N_tot} to \ref{N_sat} is flexible enough to satisfy our requirement.
The exploration of the parameter space for the HOD model is performed using the Markov Chain Monte Carlo (MCMC) method.
We use the analytic model described in \citet{Tinker_analytical, Tinker_2012} to calculate $w_{p}$ from a given HOD model. We note that the eBOSS+BOSS galaxy sample is not an ideal sample for HOD analysis. There are gaps in color space between the selection functions for each sample, thus this sample is not `complete' as is usually assumed in the standard HOD formalism. However, adding BOSS galaxies to eBOSS makes the sample significantly more complete than it would otherwise be. The bright limit on eBOSS target selection implies that the most massive halos are not represented in the sample, and the mean number of galaxies per halo cannot be assumed to monotonically increase. Inclusion of the BOSS sample brings these halos back into the fold, and meets the assumptions inherent in Eq. \ref{N_cen}.

The standard HOD approach, parameterized in Eq. \ref{N_cen}, assumes a smooth transition between halos that are not massive enough to contain a galaxy in the sample, and more massive halos that always have at least one galaxy within them. The width of this transition is determined by $\sigma_{\log{M}}$. This model naturally assumes that the sample of galaxies being modeled is complete---all galaxies above a threshold in luminosity or stellar mass are included. However, this standard approach has been used in many previous analyses of LRG samples (e.g., \citealt{wake_etal:08,Zheng_2009,CMASS_Martin,Parejko_LOWZ,Reid_2014}). In these papers, the incompleteness is folded into this transition as an extra source of scatter. The measurements of incompleteness in the BOSS samples list above make it possible to test this assumption explicitly, which we show in Appendix D. Although the impact of incompleteness on halo occupation in eBOSS+BOSS may be quantitatively different than in BOSS itself, it is possible to use the standard approach to construct an HOD model that reproduces the bias and number of density of an LRG sample with realistic incompleteness. Additionally, Figure \ref{wp_split} shows that the amplitude of clustering in our sample does not depend on redshift. Thus, the incompleteness induced by the flux limit at higher redshifts does not impact the clustering, allowing us to treat the eBOSS+BOSS galaxies as a single sample.

\begin{figure*}[htbp]
\begin{center}
\includegraphics[width=14cm, height=13cm]{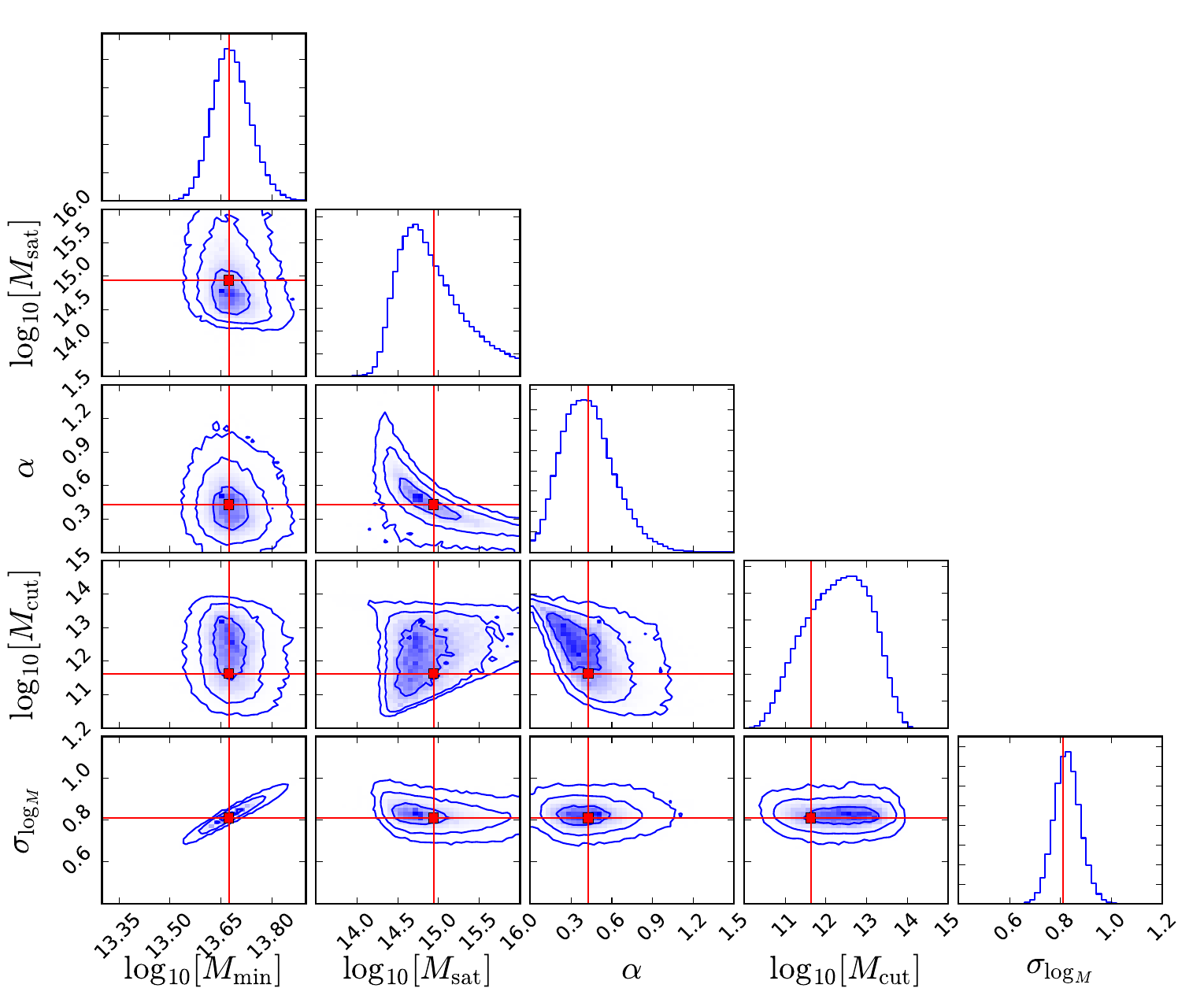}
\caption{The 68\%, 95\% and 99\% confidence intervals of the HOD parameters for the eBOSS+BOSS LRG sample based on the MCMC analysis. The diagonal panels display the one-dimensional probability distribution function. The mass parameters have the unit of $h^{-1}M_{\odot}$, and the red cross stands for the best-fit.}
\label{CL}
\end{center}
\end{figure*}

\begin{figure}[htbp]
\begin{center}
\includegraphics[width=9cm, height=10cm]{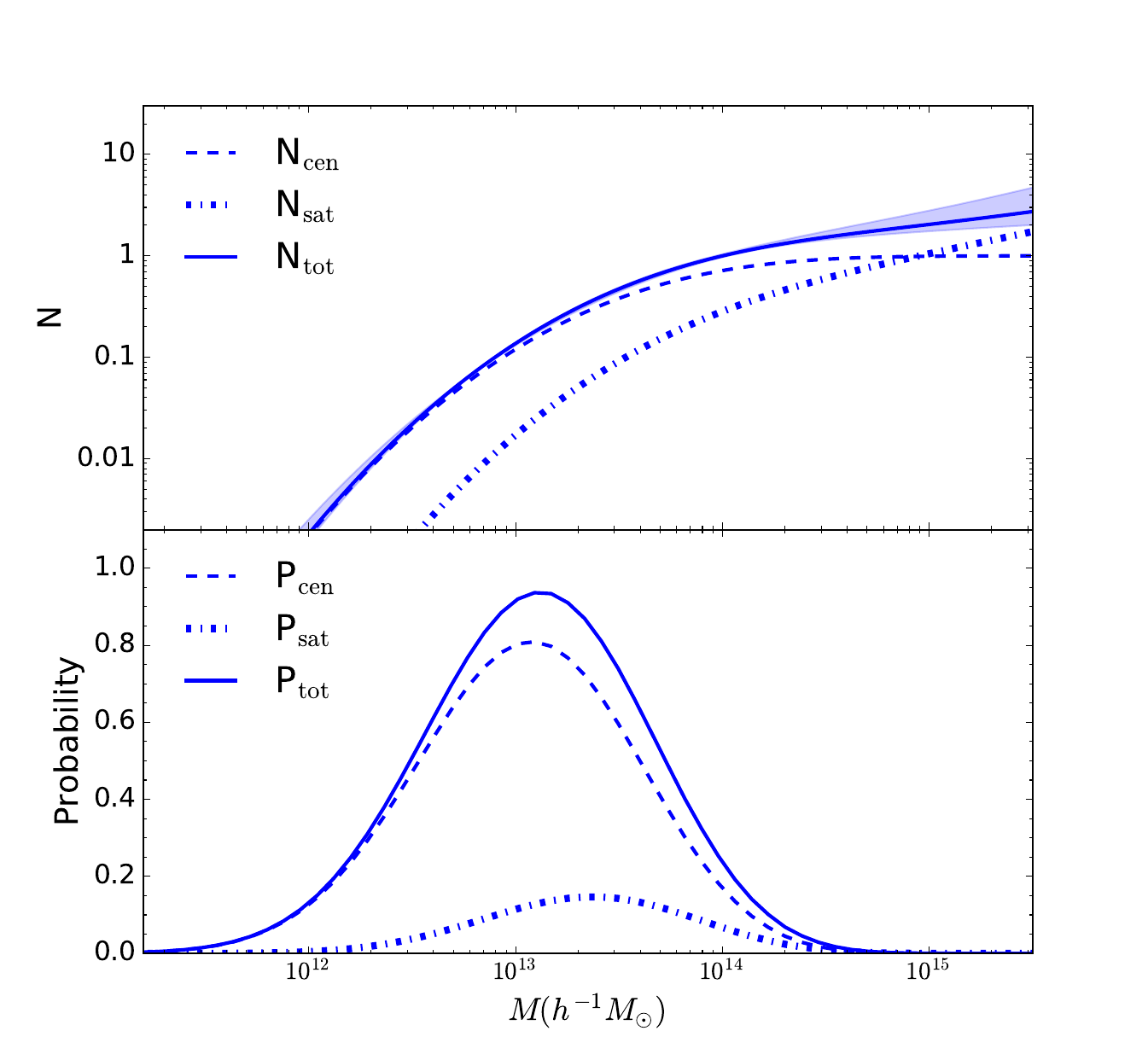}
\caption{$Top$ $panel:$ The mean occupancy of halos as a function of halo mass for the eBOSS+BOSS sample used in our calculation. The dashed, dotted and solid lines are $\rm{N}_{\text{cen}}$, $\rm{N}_{\text{sat}}$ and $\rm{N}$ respectively. The shaded regions correspond to $\pm1\sigma$ errors from the MCMC test. $Bottom$ $panel:$ Probability per $\log_{10} \rm{M}$ that a galaxy in our sample is hosted by a halo of mass $\rm{M}$. Central and satellite galaxies are shown explicitly.}
\label{HOD}
\end{center}
\end{figure}

\begin{figure}[htbp]
\begin{center}
\includegraphics[width=9cm, height=8cm]{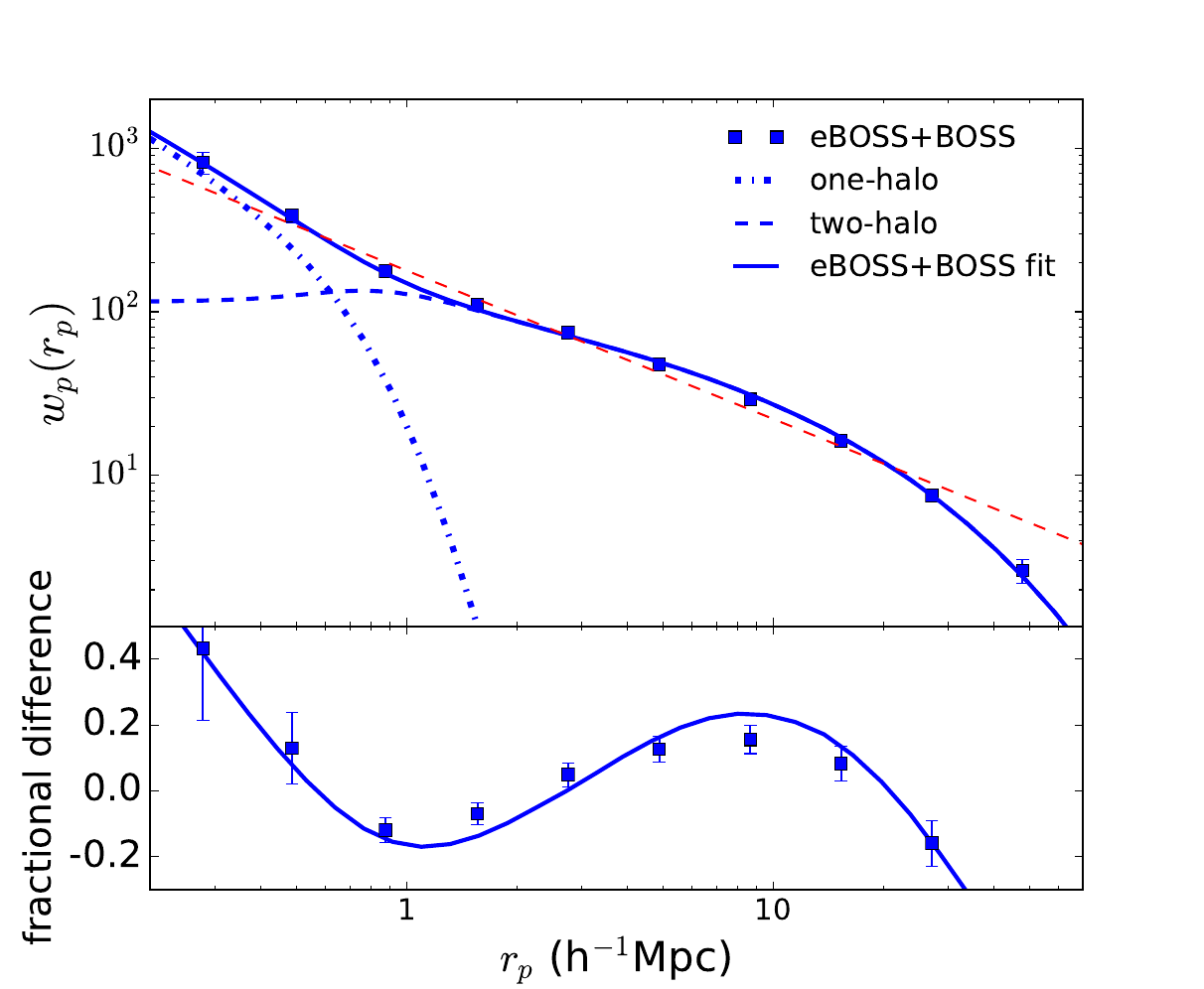}
\caption{$Top$ $panel:$ The best-fit of $w_{p}(r_{p})$ from MCMC for eBOSS+BOSS LRG samples, the one-halo term (dash-dotted) and two-halo term (dashed) are shown separately. The red dashed line is the best-fit power-law function. $Bottom$ $panel:$ The fractional difference of the clustering data and best-fit with respect to the power-law function. The best-fit $\chi^2$ is 13.6 for $\rm{N}_{\rm{d}}-\rm{N}_{\rm{p}}=6$ degrees of freedom.}
\label{wp_fit}
\end{center}
\end{figure}

\begin{table}
\caption{$Top$ 5 $rows:$ The mean and standard deviation of the HOD parameters from the Markov chain analysis. $Bottom$ 3 $rows$: $\chi^{2}$ and the derived quantities from the HOD analysis.}
\begin{center}
\begin{tabular}{rcc}
\multicolumn{1}{r}{}
 &  \multicolumn{1}{c}{eBOSS+BOSS}
 & \multicolumn{1}{c}{best-fit} \\
 \cline{1-3}
$\log{M_{\text{min}}}$ & $13.68^{+0.06}_{-0.05}$ & $13.67$ \\
$\log{M_{\text{sat}}}$ & $14.87^{+0.60}_{-0.32}$ & $14.93$ \\
$\alpha$  &  $0.41^{+0.20}_{-0.16}$  &  $0.43$ \\
$\log{M_{\text{cut}}}$ & $12.32^{+0.76}_{-0.88}$ & $11.62$\\
$\sigma_{\log{M}}$ & $0.82\pm0.05$ & $0.81$ \\
 \cline{1-3}
 $\chi^2$   & --    &  13.6 \\
$b$ & $2.30\pm0.03$ & $2.31$ \\
$f_{\text{sat}}$ & $13.0\pm3.0\%$ & $15.2\%$ \\
\cline{1-3}
\end{tabular}
\end{center}
\label{tab:HOD}
\end{table}

The constraints on the HOD parameters are presented in Figure \ref{CL} based on $\chi^2$ from the Gaussian likelihood function. 
$M_{\text{min}}$ is not a free parameter once the galaxy number density $\bar{n}$ is known and the other HOD parameters are specified. In particular, $M_{\text{min}}$ is determined by matching the number density $\bar{n}$ to the integral
\begin{equation}
\bar{n}=\int \frac{dn}{dM}N(M),
\end{equation}
where $dn/dM$ is the halo mass function from \cite{Tinker_2008}. The dark matter halo is described by the NFW profile (\citealt{NFW_1996}), and the concentration mass relation is adopted from \cite{Maccio_2008}. The values of the HOD parameters and the statistics with their confidence intervals are given in Table \ref{tab:HOD}.
These measurements are obtained by the use of the galaxy number density $\bar{n}=(1.4\pm0.05)\times10^{-4} (h^{-1} \text{Mpc})^{-3}$, which corresponds to the space density at $z=0.7$ for eBOSS+BOSS sample. We test to make sure that the characteristics of HOD are not sensitive to our choice of $\bar{n}$, the definition of characteristics refers to the galaxy bias, satellite fraction and HOD shape (which means the $M_{\rm{sat}}/M_{\rm{min}}$ ratio and $\alpha$). We repeat the analysis for the use of the maximal space density which is about twice the current one. The constraints of the HOD parameters necessarily change to account for the different numbers of galaxies, however, the characteristics of the HOD are not sensitive to the choice of the number density; the satellite fraction increases by $1\%$, while the bias remains the same, and the slope parameter $\alpha$ changes within the error estimated. This means that the HOD shape is the same.

The top panel of Figure \ref{HOD}  shows the mean occupation function of the best fit model and its uncertainties from the MCMC analysis. The mean halo mass for eBOSS+BOSS sample is $2.5\times 10^{13} h^{-1} M_{\odot}$, which is roughly in agreement with the CMASS result (\citealt{CMASS_Martin}). The bottom panel displays the probability that a galaxy in our sample is hosted by a halo of mass $M$. The galaxies observed in the survey live in a wide halo mass distribution. 

The best-fit of the $w_{p}$ from the HOD modeling is presented in Figure \ref{wp_fit}, where the one-halo term and two-halo term are also shown for illustration\footnote{The one-halo term means that the two galaxies in the pair come from the same halo, while the two-halo term means they come from two distinct halos.}. The transition scale from one-halo term to two-halo term is observed at $\sim$1-2 $h^{-1}$ Mpc.

\begin{figure}[htbp]
\begin{center}
\includegraphics[width=9cm, height=7cm]{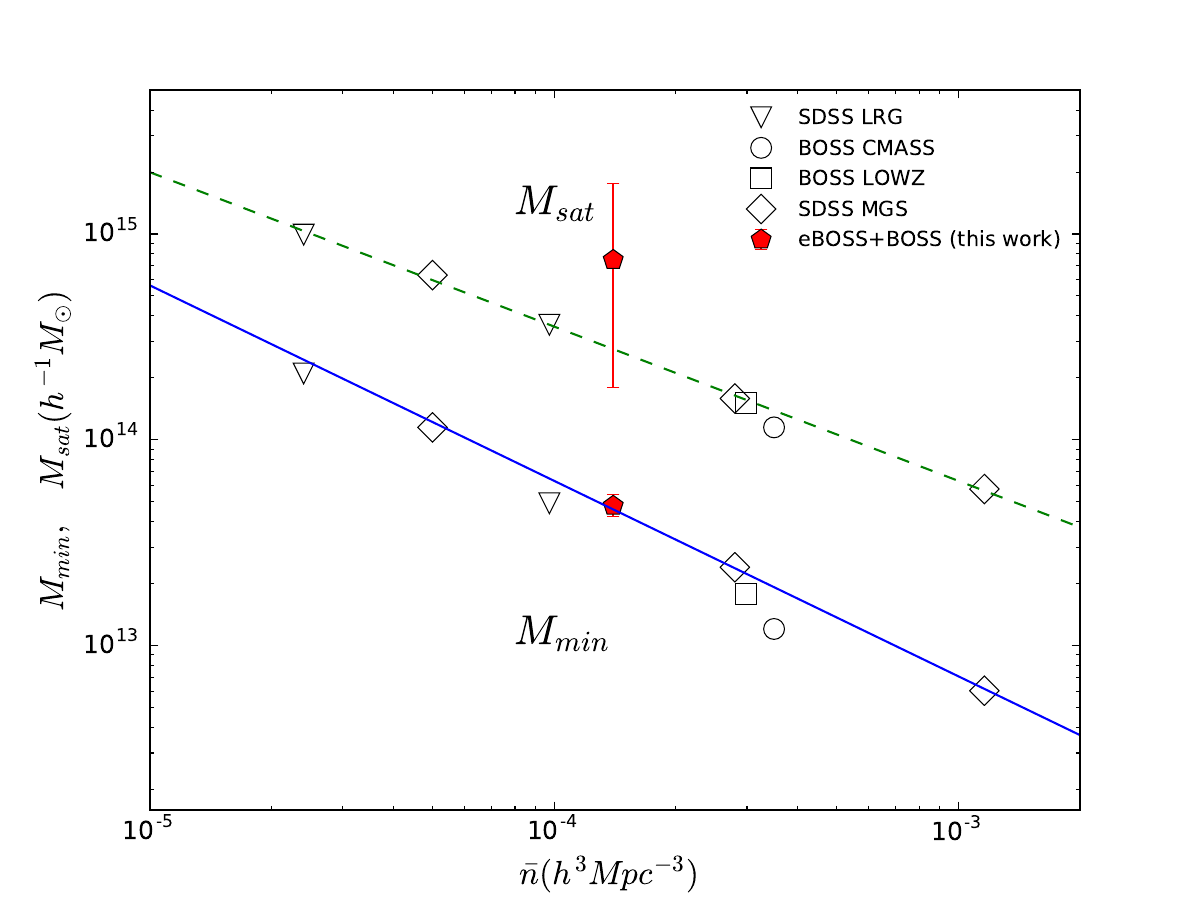}
\caption{The HOD parameters $M_{\text{sat}}$ and $M_{\text{min}}$ as a function of galaxy number density for different galaxy samples. The error bars are not shown for clarity except for the eBOSS+BOSS results. The labels refer to following studies: SDSS LRG: \citet{Zheng_2009}, BOSS CMASS: \citet{CMASS_Martin}, BOSS LOWZ: \citet{Parejko_LOWZ}, SDSS MGS: \citet{Zehavi_2011}, eBOSS and eBOSS+BOSS: this work. The solid and dashed lines roughly show the linear tendency of $M_{\text{min}}$ and $M_{\text{sat}}$ with respect to the number density.}
\label{mass}
\end{center}
\end{figure}

The HOD modeling of massive galaxies at different redshifts has been investigated with various samples and HOD models. We compare our measurements of HOD parameters $M_{\rm{min}}$ and $M_{\rm{sat}}$ versus galaxy number density $\bar{n}$, with other studies which use the similar statistical method, in Figure \ref{mass}; these include the samples of SDSS (\citealt{Zehavi_2011}), BOSS CMASS (\citealt{CMASS_Martin}), and BOSS LOWZ (\citealt{Parejko_LOWZ}).
Our HOD fitting results are in reasonable agreement with those of previous studies. 
The value of $M_{\text{sat}}$ from our eBOSS+BOSS measurement appears to be somewhat above the trend. A larger satellite mass scale would normally imply a smaller fraction of satellites, but the $f_{\text{sat}}$ value from eBOSS+BOSS is in good agreement with CMASS and LOWZ results, all near $~10\%$. For these BOSS results, as well as the SDSS results, $\alpha$ is near unity, while our best-fit value is 0.43. There is a strong degeneracy between $\alpha$ and $M_{\text{sat}}$ (see Figure \ref{CL}), such that a value of $\alpha\sim1$ from the eBOSS+BOSS sample would bring $M_{\text{sat}}$ into better agreement with the other surveys.

\cite{Zehavi_2011} estimate that $M_{\text{sat}}\approx 17M_{\text{min}}$ in the SDSS galaxy sample. Incorporating the mass estimates presented in Figure \ref{mass}, we find this relationship depends on number density as $M_{\text{sat}}\approx 17M_{\text{min}}(\bar{n}/\bar{n}_{\text{SDSS}})^{0.2}$, where $\bar{n}_{\text{SDSS}}$ is the number density of SDSS galaxy samples. This result implies that in the low space density environment, the gap between the masses of the halos which host two galaxies and the one hosting only one galaxy is smaller than in the dense environment. 

Based on the HOD fitting, we estimate the bias of the galaxy sample with respect to the dark matter distribution through
\begin{equation}
b = \bar{n}^{-1}\int_{0}^{\infty}b_{h}(M)N(M)\frac{dn}{dM}dM,
\end{equation}
where $b_{h}(M)$ is the halo bias factor from \cite{Tinker_2010}, and there is no radial range in which to measure the galaxy bias.
The large scale bias of eBOSS+BOSS sample is $2.30\pm0.03$ from our clustering measurements. This value varies inversely with the assumed mass perturbation amplitude $\sigma_{8}$, which is set to be 0.8 in this work.

As a consistency check, we have also determined the galaxy bias independently by simply taking the ratio of the measured projected correlation and the theoretical linear dark matter projected correlation. Here we considered only points well within the linear regime ($r_p > $ 3 Mpc$/$h). This method is independent of the HOD modeling and fit details, and yet produces a consistent measurement of bias $b=2.34 \pm 0.02$, which is reassuring of our methodology. 
 
We note that the high $\chi^2$ of the best-fit model, 13.6, is driven by relatively poor agreement with the data at $1<r_p<3$ $h^{-1} \text{Mpc}$. This is likely a failure of the scale-dependent bias model used, which is calibrated on lower-mass halos and lower-redshift samples, and is the chief uncertainty in HOD fitting (see, e.g., \citealt{Tinker_2012}). A more flexible HOD model with more freedom in the modeling of the scale-dependent bias may yield a lower $\chi^2$, but the characteristics of the galaxy sample \textemdash the bias and $f_{sat}$ \textemdash are unlikely to change. In tests we find that ad-hoc changes to the scale-dependent bias formula do lower the $\chi^2$ of fit, but the characteristics of the HOD itself do not change outside of our $1\sigma$ statistical errors.

\subsection{Redshift Evolution}

\begin{figure}[htbp]
\begin{center}
\includegraphics[width=9.5cm, height=9cm]{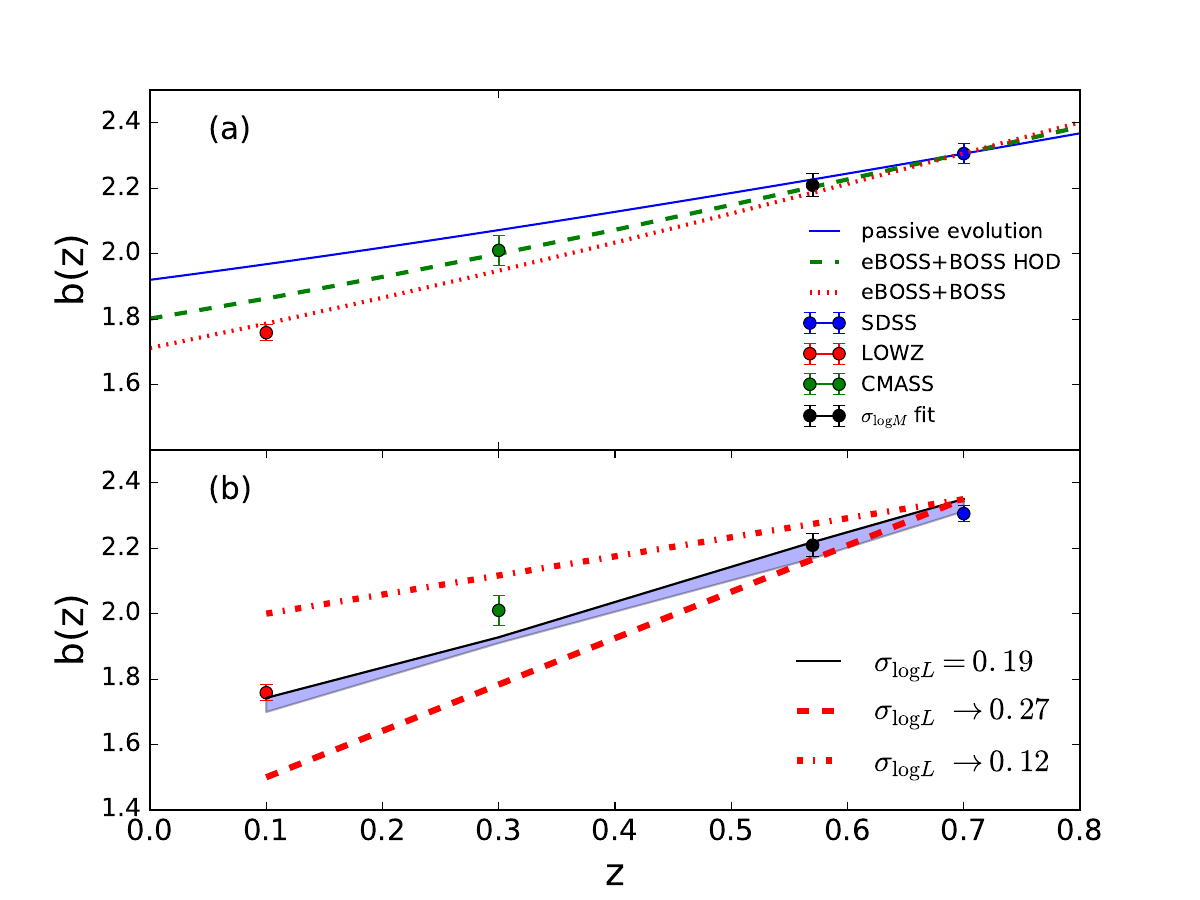}
\caption{Measurements of bias based on the clustering studies. $Top$ $panel:$ Dynamical passive evolution model from eBOSS+BOSS LRGs (blue solid line). The green dashed line is the bias produced by placing the eBOSS+BOSS HOD at various redshifts. The red dotted line is obtained from the fit of the parameter $\sigma_{\log{M}}$ as explained in the context. $Bottom$ $panel:$ The interpretation of a constant scatter $\sigma_{\log{L}}=0.19$ of these bias measurements (Solid). The blue shaded region corresponds to $\sigma_{\log{L}}$ between 0.19 and 0.20. Dashed and dot-dashed lines correspond to a linear evolution of $\sigma_{\log{L}}$ as a function of redshift $z$ from $\sigma_{\log{L}}=0.19$ at $z=0.7$ to $\sigma_{\log{L}}=0.27$ and $\sigma_{\log{L}}=0.12$ at $z=0.1$, respectively.}
\label{bias}
\end{center}
\end{figure}

The bias from the eBOSS+BOSS sample is significantly larger than the BOSS results of  \citet{CMASS_Martin} and \citet{Parejko_LOWZ}. 
To make a robust comparison between various samples, we compare our eBOSS bias measurement to galaxy samples from BOSS CMASS (\citealt{CMASS_Martin}, $z\sim 0.57$), BOSS LOWZ (\citealt{Parejko_LOWZ}, $z\sim0.3$) and the SDSS Main Galaxy Sample (\citealt{Zehavi_2011}, $z\sim0.1$) at fixed number density.
For each sample, we rank-order the galaxies by absolute magnitude and truncate the sample at the magnitude limit that achieves a space density of $\bar{n}_{g}=1.4\times10^{-4} (h^{-1} \text{Mpc})^{-3}$. For the CMASS and LOWZ samples, we also restrict the redshift range of the samples to be $\Delta z = \pm0.1$ around the median redshift. 
This minimizes the incompleteness of theses samples, see further details in the Appendix \ref{subsample} and Figure \ref{wp_sub}. For CMASS and LOWZ, this procedure retains 50\% of the samples. For the SDSS Main Galaxy Sample, we create a volume-limited sample with $\text{M}_{r}<-21.7$ galaxies to obtain the same number density. For each sample, we measure the new bias as described more detailedly in the appendix. The new bias values are shown as a function of redshift in Figure \ref{bias}. When fixing  $\bar{n}_{g}$, the $b(z)$ results show a fairly linear trend with redshift, opposed to the full-sample BOSS analysis that found $b\sim2.0$ for both samples (\citealt{CMASS_Martin, Parejko_LOWZ}).

The top panel of Figure \ref{bias} compares these data to the prediction of the passive evolution model \citep{Fry_bias}, which significantly underpredicts the evolution of bias with redshift. \cite{Guo_2013} compare the passive evolution model with the clustering of CMASS galaxies and find consistent result, but in a narrower redshift range of $0.47<z<0.62$. Using a much larger redshift range as Figure \ref{bias}a, it highlights the deficiency of the passive evolution model to describe the clustering of massive galaxies.
We also show a model in which the best-fit HOD from eBOSS+BOSS is used to predict the bias at the median redshift of each survey. This ansatz --- that halo occupation of massive galaxies does not evolve --- predicts more evolution in $b(z)$ than the passive model, but is still not a good description of the data especially at low redshift. Therefore some evolution is required for the full description from $z=0.7$ to $z=0$. The amount of bias evolution in the data implies that the HOD is evolving with time; namely, the scatter parameter $\sigma_{\log{M}}$ must increase with cosmic time to lower the bias at lower $z$. The dotted curve in top panel of Figure \ref{bias}  shows a model in which the scatter in halo mass at fixed luminosity varies with redshift as 
\begin{equation}
\sigma_{\log{M}}(z)=\sigma_{\log{M}}(z=0.7)\left(\frac{1+z}{1+0.7}\right)^{\beta}
\end{equation}
with $\beta\sim-0.3$, which yields a nearly linear fit to the $b(z)$ measurements. Although neither model is perfect, the fixed eBOSS+BOSS HOD model (green dashed line) yields a $\chi^{2}=19.9$ compared with the data, while the the evolving $\sigma_{\log{M}}$ model (red dotted line) yields a $\chi^{2}=8.7$ which is preferred by the data.

Just because the scatter in halo mass at fixed luminosity varies with redshift does not necessarily imply that the scatter in luminosity at fixed halo mass ($\sigma_{\log{L}}$, hereafter) is also changing. To convert from one scatter to another requires the logarithmic slope of the halo mass function, which is also evolving with time. For galaxy formation theory, $\sigma_{\log{L}}$ is the more fundamental parameter, as it indicates how formation efficiency can vary at fixed gravitational potential, see e.g. \citet{Gu_2016}. The lower panel in Figure \ref{bias} compares our $b(z)$ data to predictions from the abundance-matching model (see, e.g. \citealt{Behroozi_2013} and references therein). Here, we adopt the $r$-band luminosity function measured from the AGES survey \citet{Cool_2012} to match galaxy luminosity onto halo mass. We use the high-resolution MultiDark N-body simulation presented in \citet{Riebe_2011, Behroozi_2013b, Behroozi_2013}, as well as the method presented in \citet{Wetzel_Martin_2010} to incorporate scatter at fixed halo mass. The data are consistent with a redshift-independent scatter of $\sigma_{\log{L}}=0.19$, thus the change in $\sigma_{\log{M}}$ is entirely due to the evolution in the halo mass function and not due a change in the growth of stellar mass in massive objects over time. Moreover, Figure \ref{bias}b highlights just how sensitive these data are to $\sigma_{\log{L}}$; the other curves show models in which $b(0.1)=2.0$ --- i.e., scatter is shrinking --- and $b(0.1)=1.5$, in which scatter increases with time. Our measurements imply $\sigma_{\log{L}}=0.19\pm0.02$ with no redshift evolution.

\section{Discussion and Conclusion}

This paper marks the first scientific results from the eBOSS LRG program. Although the observing strategy for eBOSS LRGs differs substantially from its predecessors in BOSS and SDSS, we have demonstrated that the combination of the bright-end of the BOSS CMASS sample with the eBOSS LRGs over the redshift range $0.6<z<0.9$ provides a robust clustering sample at small and intermediate scales. Our halo occupation analysis of this sample indicates that these galaxies have properties that are well-placed within our understanding of the relationship between massive galaxies and dark matter halos, with a bias factor of $b=2.30$, a satellite fraction of $\sim 13\%$, and halo mass scale in agreement with the scaling relations calibrated on other surveys. The addition of the eBOSS galaxy sample to previous spectroscopic samples yields a set of massive galaxies that span that last $\sim 7$ Gyr of the history of the universe. 

Our measurement of scatter in galaxy luminosity at fixed halo mass, $\sigma_{\log{L}}=0.19\pm0.02$, is in good agreement with other studies that have focused on $z=0$ samples. \cite{lehmann_etal:15}, using galaxy clustering alone, reported a value of $0.17^{+0.03}_{-0.05}$; \cite{reddick_etal:13}, using a combination of galaxy groups and clustering, find $0.21^{+0.01}_{-0.02}$; and \cite{more_etal:09}, using satellite kinematics, find $0.16\pm0.04$. Assuming these measurements are all independent (which is not strictly true), the weighted combination of all four results indicate $\sigma_{\log{L}}=0.19\pm 0.01$, a value that is somewhat larger than recent measurements of the scatter in stellar mass at fixed halo mass, $\sigma_{\log{M\ast}}\approx 0.16$ (\citealt{li_etal:12, kravtsov_etal:14, tinker_etal:16_boss, zu_mandelbaum:16}), which itself appears to be independent of redshift. The larger scatter in luminosity, for galaxies that are nearly all on the red sequence, is indicative of different formation histories at fixed stellar mass that yield different stellar-$M/L$ ratios and mean stellar ages. 

At first glance, the lack of evolution of either scatter value is notable but not surprising given that the massive end of the red sequence is constructed prior to $z\sim 1$ and that massive galaxies evolve in a manner close to passive stellar evolution over that timespan (\citealt{wake_etal:08, cool_etal:08}). However, true passive evolution of massive galaxies would result in a reduction in $\sigma_{\log{L}}$ as galaxies evolve, due to the fact that $M/L$ ratios for passive stellar populations evolve to the same asymptotic value. To match dynamically passive evolution, $\sigma_{\log{L}}$ would have to decrease from 0.19 at $z=0.7$ to 0.12 at $z=0.1$, which is clearly ruled out by our measurements. \cite{Gu_2016} find that the scatter (in stellar mass) induced by hierarchical merging is constant with redshift, but merging is not the dominant source of scatter at the halo masses probed by eBOSS galaxies. For galaxies in $10^{13}$ $h^{-1}M_{\odot}$ halos, in-situ star formation is still predicted to be the dominant source of scatter. Abundance-matching studies by \cite{Behroozi_2013b} and \cite{moster_etal:13} demenstrate that stellar mass growth from merging accounts for $\sim 10\%$ of the $z=0$ galaxy mass. This result is in agreement with earlier clustering studies of massive galaxies that found LRG merger rates of $\sim 1\%$ per Gyr (\citealt{wake_etal:08} and references therein). How does a population without merging or star formation have a constant luminosity scatter for over half the lifetime of the universe? 

SDSS, CMASS, LOWZ, and eBOSS represent a heterogeneous set of galaxy samples. Our SDSS sample is volume-limited, and at $M_r<-21.7$ the fraction of star-forming objects is negligible. The BOSS samples, as a whole, suffer from high significant incompleteness due to their color-based selections (\citealt{leauthaud_etal:16, tinker_etal:16_boss}), but by using only the brightest third of each sample in relatively narrow redshift ranges, CMASS and LOWZ are roughly complete as well. eBOSS, however, cannot be considered a complete sample. It is not trivial to estimate what the bias of a complete eBOSS sample would be at the number density used to create our subsamples, $1.4\times 10^{-4}(h^{-1} \text{Mpc})^{-3}$. The color selection excludes some brighter galaxies and includes some fainter objects, but the fainter objects will be redder and thus possibly more clustered than the brighter, but bluer, excluded objects. This is true of the overall CMASS sample (c.f. Figure 7 of \citealt{tinker_etal:16_boss}). If this is true of eBOSS, then the overall trend of $b(z)$ in Figure \ref{bias} would be consistent with some small reduction in $\sigma_{\log L}$ with time. Alternatively, the scatter in stellar $M/L$-ratio on the red sequence may not change enough between $z=0.7$ and $z=0.1$ to be detectable within our precision of 0.02 dex in scatter, since this scatter would add in quadrature with the scatter in stellar mass at fixed halo mass. Stellar population synthesis models would be required to address this question within the precision of our measurements, and will be included in a future work.

The primary science driver of the eBOSS LRG sample is to probe the growth and expansion history of the universe at $z=0.7$. As a part of the SDSS-IV project, the eBOSS survey takes over the mission from its precursor BOSS and will map the universe in a higher redshift range and larger volume. After roughly one year observation, we reach a LRG sample with more than 34000 massive galaxies at an effective redshift $z\sim 0.7$. The result here shows that eBOSS is working well and the designed expectation is being reached. The clustering measurements that will be achieved with this sample through the completion of this survey will an important extension toward a complete map of the observable universe.




\acknowledgements{}

We thank Peter Behroozi for having his Rockstar DM halo catalogs publicly available, as well as Pierre Laurent for his kind help. Z.Z., J.L.T., and M.R.B. are supported by NSF grant F8670. H.-J. Seo is supported by the U.S. Department of Energy, Office of Science, Office of High Energy Physics under Award Number
DE-SC0014329.

Funding for the Sloan Digital Sky Survey IV has been provided by
the Alfred P. Sloan Foundation, the U.S. Department of Energy Office of
Science, and the Participating Institutions. SDSS-IV acknowledges
support and resources from the Center for High-Performance Computing at
the University of Utah. The SDSS web site is www.sdss.org.

SDSS-IV is managed by the Astrophysical Research Consortium for the 
Participating Institutions of the SDSS Collaboration including the 
Brazilian Participation Group, the Carnegie Institution for Science, 
Carnegie Mellon University, the Chilean Participation Group, the French Participation Group, Harvard-Smithsonian Center for Astrophysics, 
Instituto de Astrof\'isica de Canarias, The Johns Hopkins University, 
Kavli Institute for the Physics and Mathematics of the Universe (IPMU) / 
University of Tokyo, Lawrence Berkeley National Laboratory, 
Leibniz Institut f\"ur Astrophysik Potsdam (AIP),  
Max-Planck-Institut f\"ur Astronomie (MPIA Heidelberg), 
Max-Planck-Institut f\"ur Astrophysik (MPA Garching), 
Max-Planck-Institut f\"ur Extraterrestrische Physik (MPE), 
National Astronomical Observatory of China, New Mexico State University, 
New York University, University of Notre Dame, 
Observat\'ario Nacional / MCTI, The Ohio State University, 
Pennsylvania State University, Shanghai Astronomical Observatory, 
United Kingdom Participation Group,
Universidad Nacional Aut\'onoma de M\'exico, University of Arizona, 
University of Colorado Boulder, University of Oxford, University of Portsmouth, 
University of Utah, University of Virginia, University of Washington, University of Wisconsin, 
Vanderbilt University, and Yale University.

\appendix

\renewcommand\thefigure{\thesection.\arabic{figure}}

\section{bright star mask}
\label{bright}
\setcounter{figure}{0}

To investigate the effect of the bright stars on the LRG clustering, we apply the latest bright star mask which is designed for eBOSS tiling process to our clustering measurements.
The bright source catalog used for the analysis is based on the WISE Allsky catalog\footnote{http://wise2.ipac.caltech.edu/docs/release/allsky/}. 
All the sources with at least one saturated pixel are selected. 

\begin{figure}[htbp]
\begin{center}
\includegraphics[width=8cm, height=6cm]{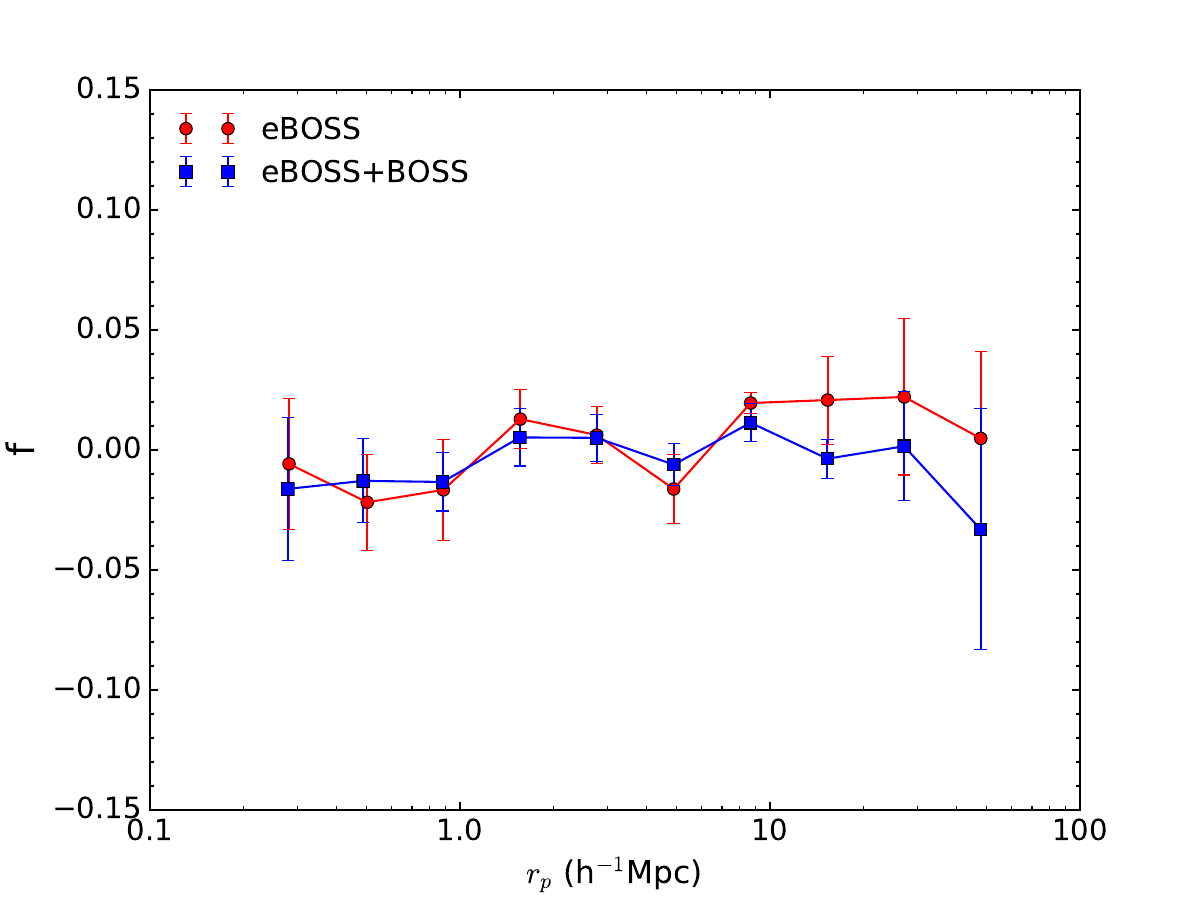}
\caption{The effect of the bright star mask on the clustering measurements. It is represented by the fractional difference of the projected correlation $w_{p}$ with and without applying the bright star mask. Both the eBOSS and eBOSS+BOSS LRG samples are shown, and the effect is found to be smaller than 5\% at all scales. }
\label{f_wp}
\end{center}
\end{figure}

In order to present the influence of the bright star mask, we calculate the following quantity
\begin{equation}
f(r_{p})=\frac{w_{p2}(r_{p})-w_{p1}(r_{p})}{w_{p1}(r_{p})},
\end{equation}
where $w_{p2}$ and $w_{p1}$ are the projected correlation functions for LRG with and without applying the bright star mask respectively.
We assess the value of this fractional difference $f$ through tens of random catalogs which have different sizes and seeds. Figure \ref{f_wp} presents the average and $1\sigma$ error of $f$ from ten different realizations of random catalogs. This result shows that the effect of bright star mask is no more than $5\%$ at all scales. This deviation is therefore believed to be noise dominated and not significant. The HOD interpretation of the clustering measurement due to the bright star mask is still valid since the HOD parameters have no essential change.

\section{tiled mock test}
\label{mock}
\setcounter{figure}{0}

\begin{figure}[htbp]
\begin{center}
\includegraphics[width=8cm, height=7.5cm]{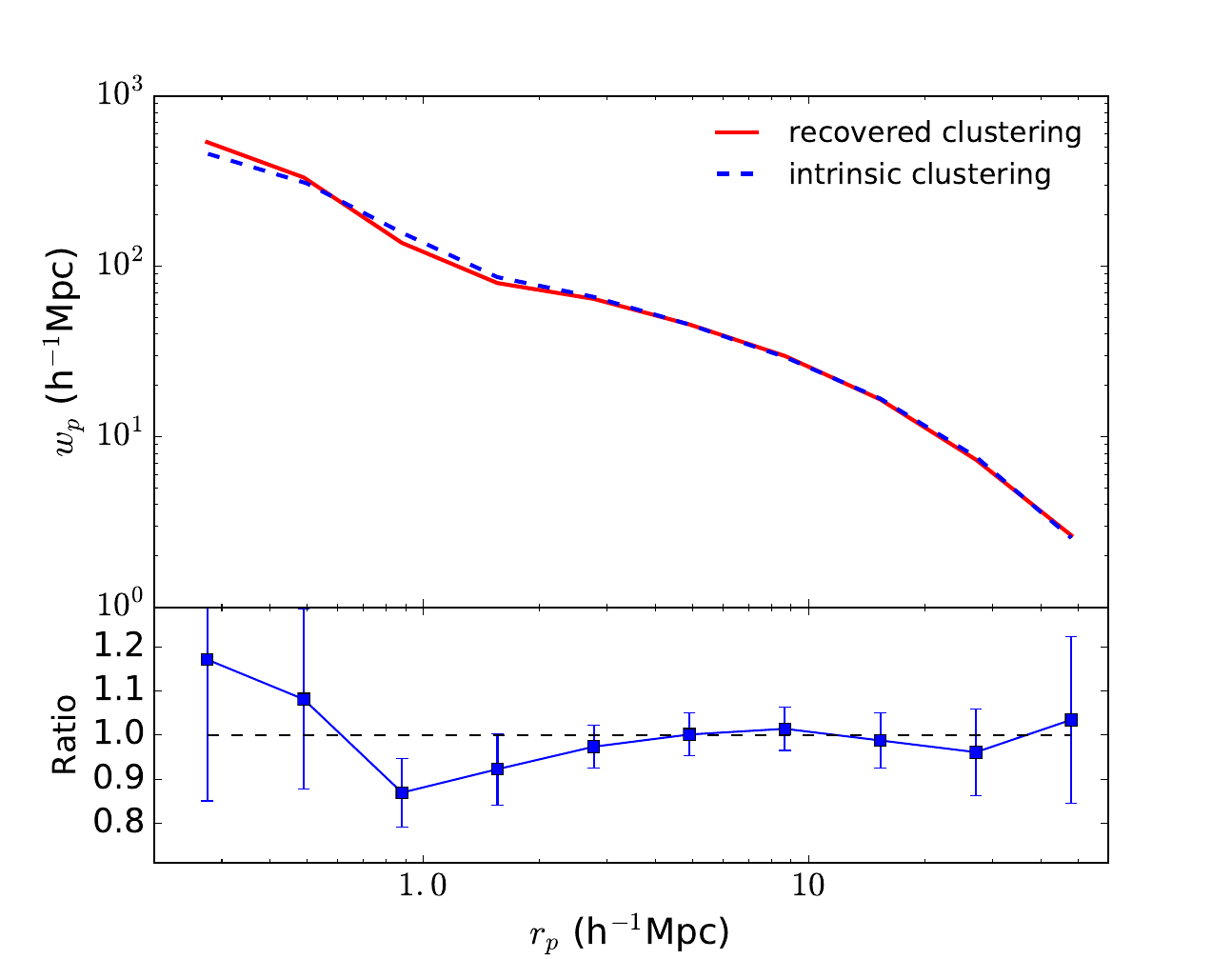}
\includegraphics[width=8cm, height=7.5cm]{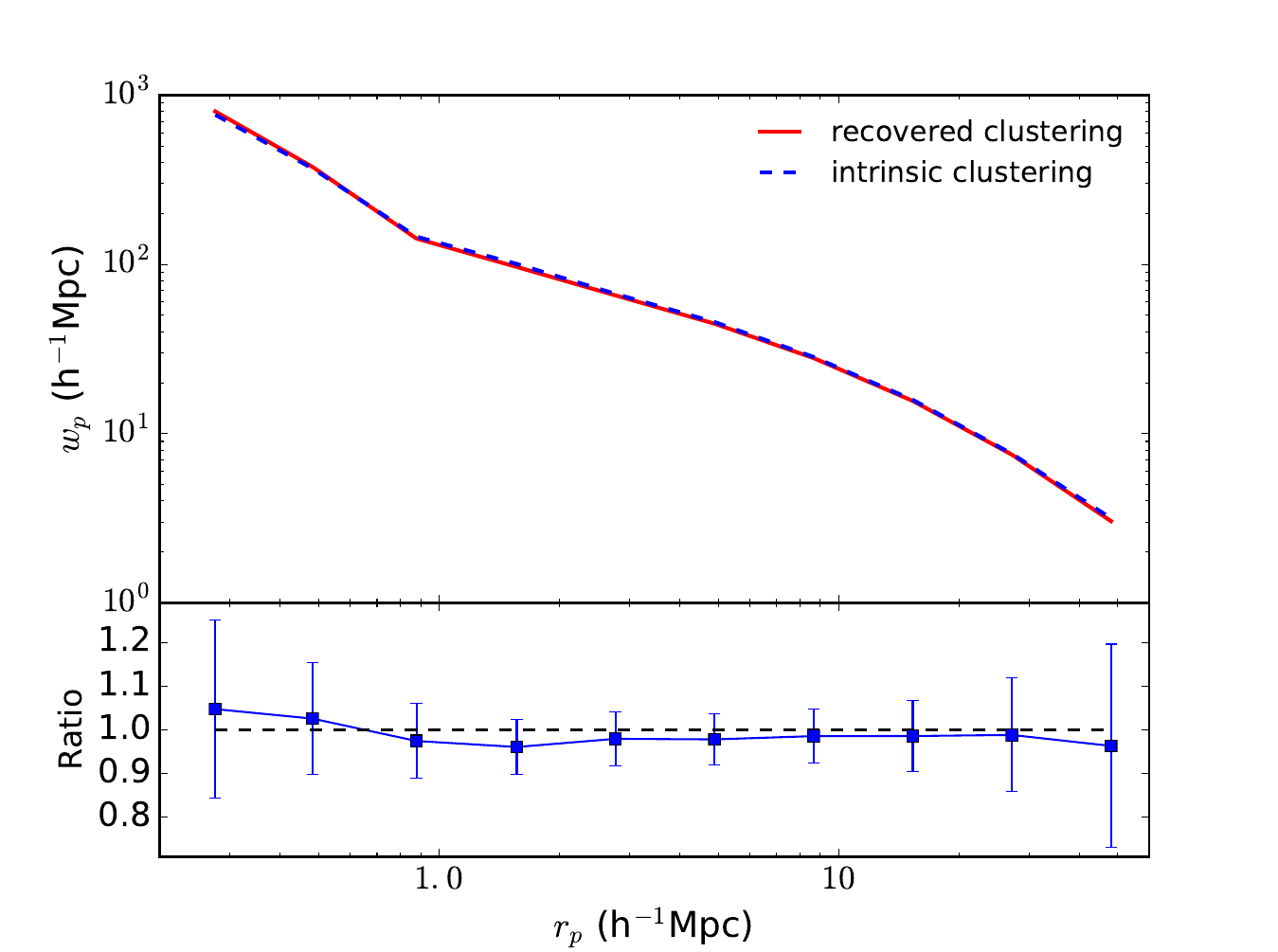}
\caption{The clustering measurement from the tiled mock. The intrinsic clustering is also shown for comparison. The recovered clustering is measured by the correction method as described in Section \ref{sec:clustering}. The bottom panel shows the ratio between the recovered clustering and intrinsic clustering. The consistency between these results especially in small scales approves our method to measure the correlation function. $Left$  $panel:$ eBOSS; $Right$ $panel:$ BOSS}
\label{mock_wp}
\end{center}
\end{figure}

The effect of the fiber allocation on our clustering measurement is tested with a mock catalog. This mock is put into the same tiling process as the eBOSS survey. The resulting catalog has the same survey geometry, redshift distribution and target density as the LRG sample. Therefore we apply the same angular correction method to this sample; the result is shown in the left panel of Figure \ref{mock_wp}. The consistency between the intrinsic clustering and the recovered clustering is clear, thus validating our measurement method. For comparison, we also test this correction method for the BOSS CMASS mock (right panel) which reveals the same robustness. The agreement is better for BOSS data than eBOSS because the corrections to the small scale pair counts are much smaller. Poisson noise is more significant for small-scale eBOSS pair counts because a significantly higher fraction --- roughly a factor of two --- of small-angle pairs are lost.

\section{Bias measurements for the CMASS, LOWZ and SDSS subsamples}
\label{subsample}
\setcounter{figure}{0}

In order to compare the measurements of bias at different redshifts, a natural choice is to construct galaxy samples with equivalent cumulative number density (\citealt{Behroozi_2013c}). Choosing number density removes much of the uncertainties in comparing galaxy samples at fixed magnitude or stellar mass thresholds, given the evolution in such quantities, as well as the uncertainties in comparing disparate samples. 
We note that comparing samples at fixed number density does not remove all possible biases (see, e.g., \citealt{Behroozi_2013c, Contreras_2017}), but it is preferable to other available alternatives. The samples we compare to the eBOSS+BOSS sample are the CMASS sample, the LOWZ sample, and the SDSS main galaxy sample (MGS). All three of these samples have higher number density than the eBOSS+BOSS sample, thus we create subsamples of each of these samples that have our sample number density of $1.4\times 10^{-4} (h^{-1} \text{Mpc})^{-3}$. For the MGS, this process is straightforward. We use the volume-limited $r$-band samples supplied as part of the NYU-VAGC (\citealt{Blanton_2005a}). The $M_r-5\log h<-21.5$ volume-limited sample has a larger number density than the eBOSS+BOSS sample. From this sample, we select all galaxies brighter than $M_r-5\log h=-21.7$ to match the desired galaxy number density.

\begin{figure}[htbp]
\begin{center}
\includegraphics[width=5.5cm, height=5.0cm]{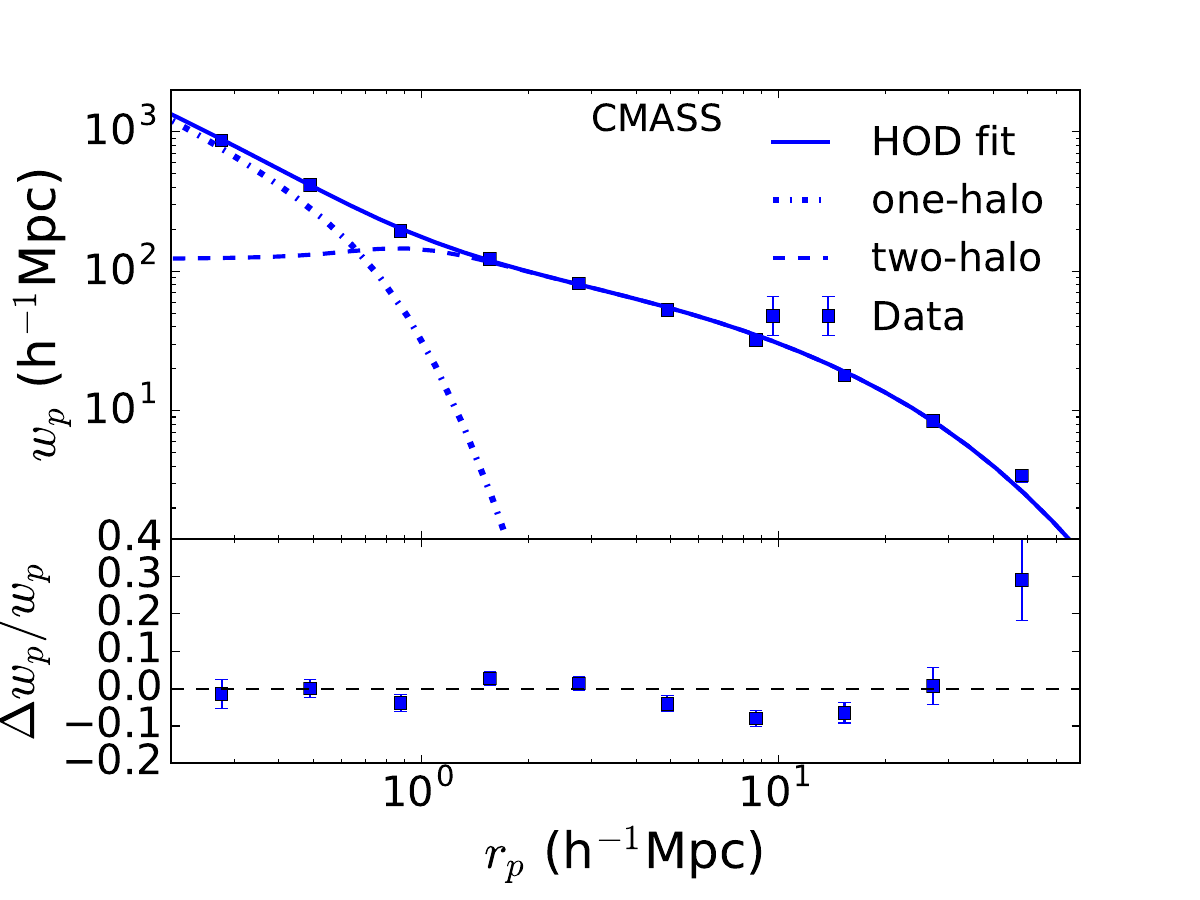}
\includegraphics[width=5.5cm, height=5.0cm]{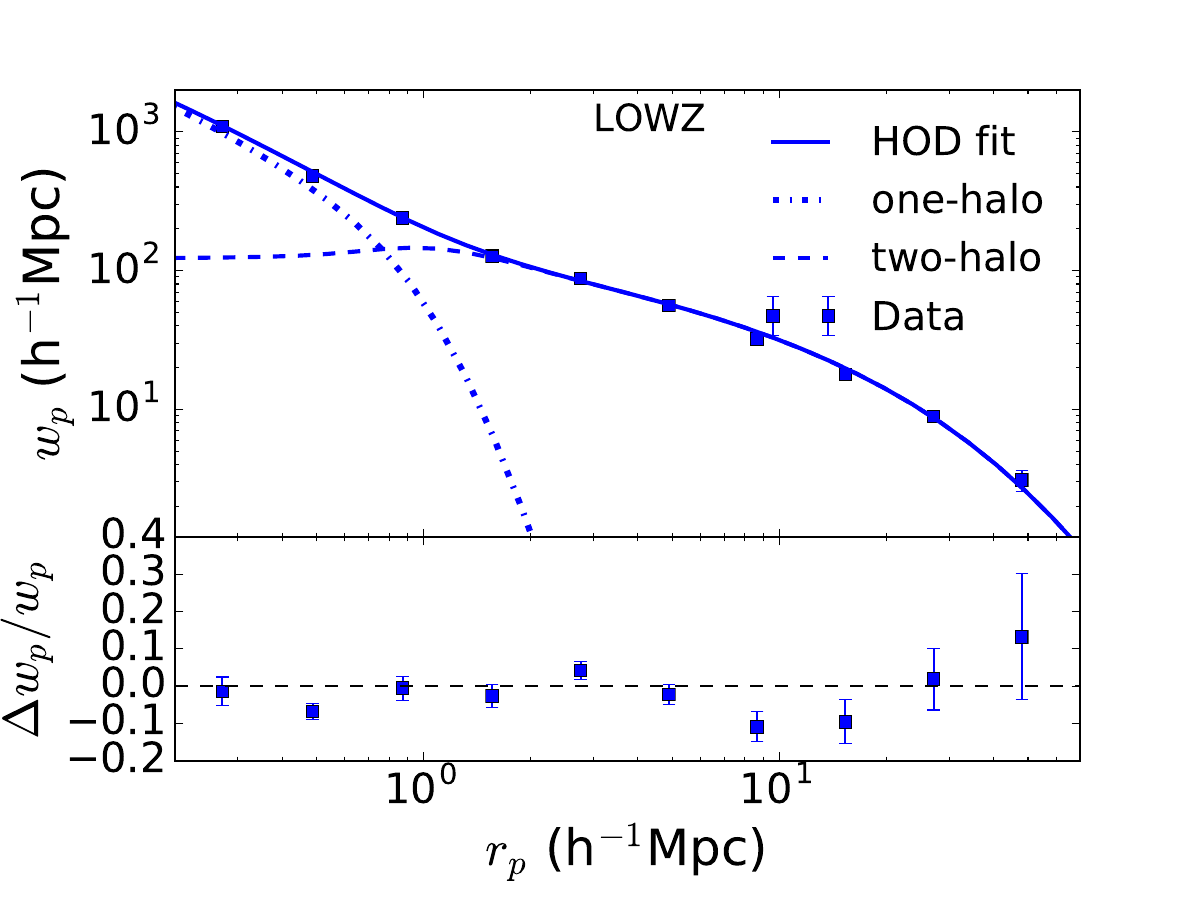}
\includegraphics[width=5.5cm, height=5.0cm]{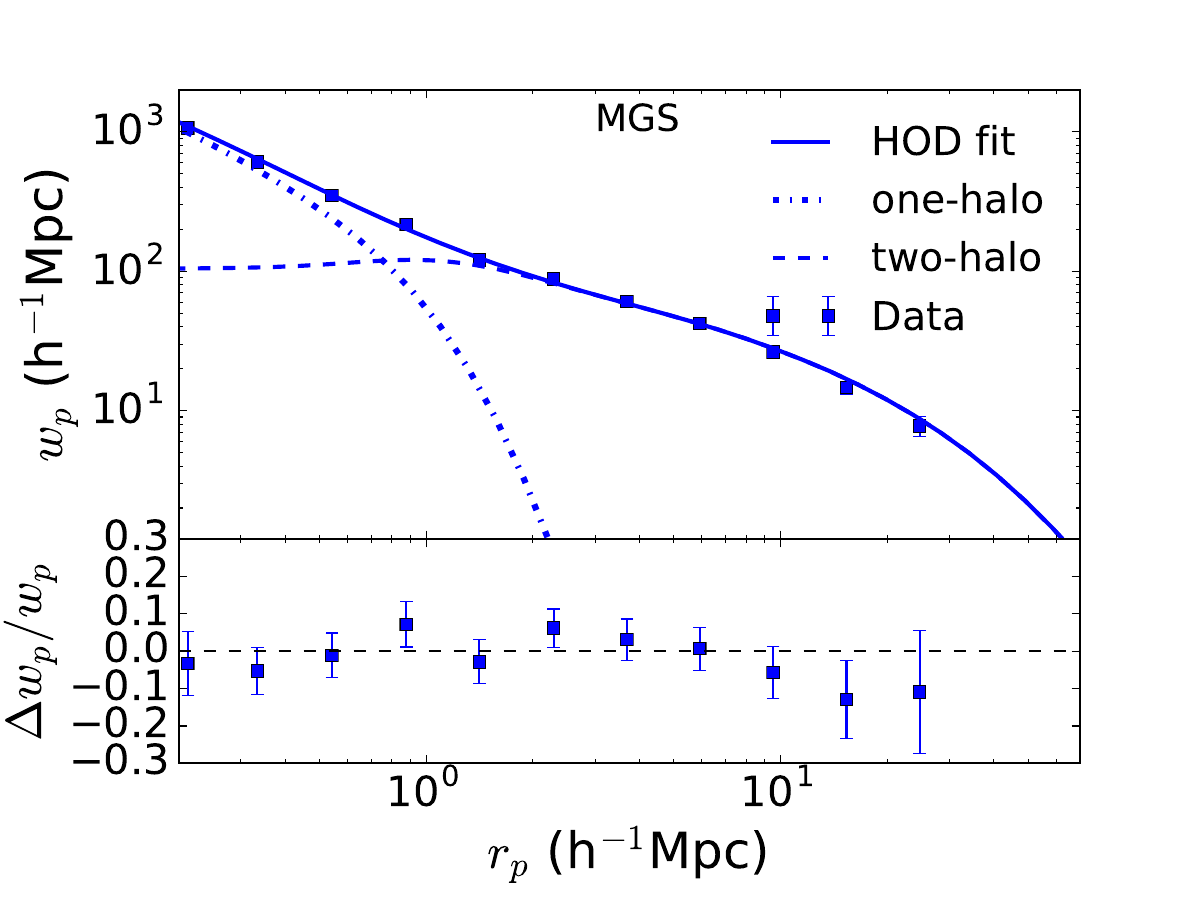}
\includegraphics[width=5.5cm, height=5.0cm]{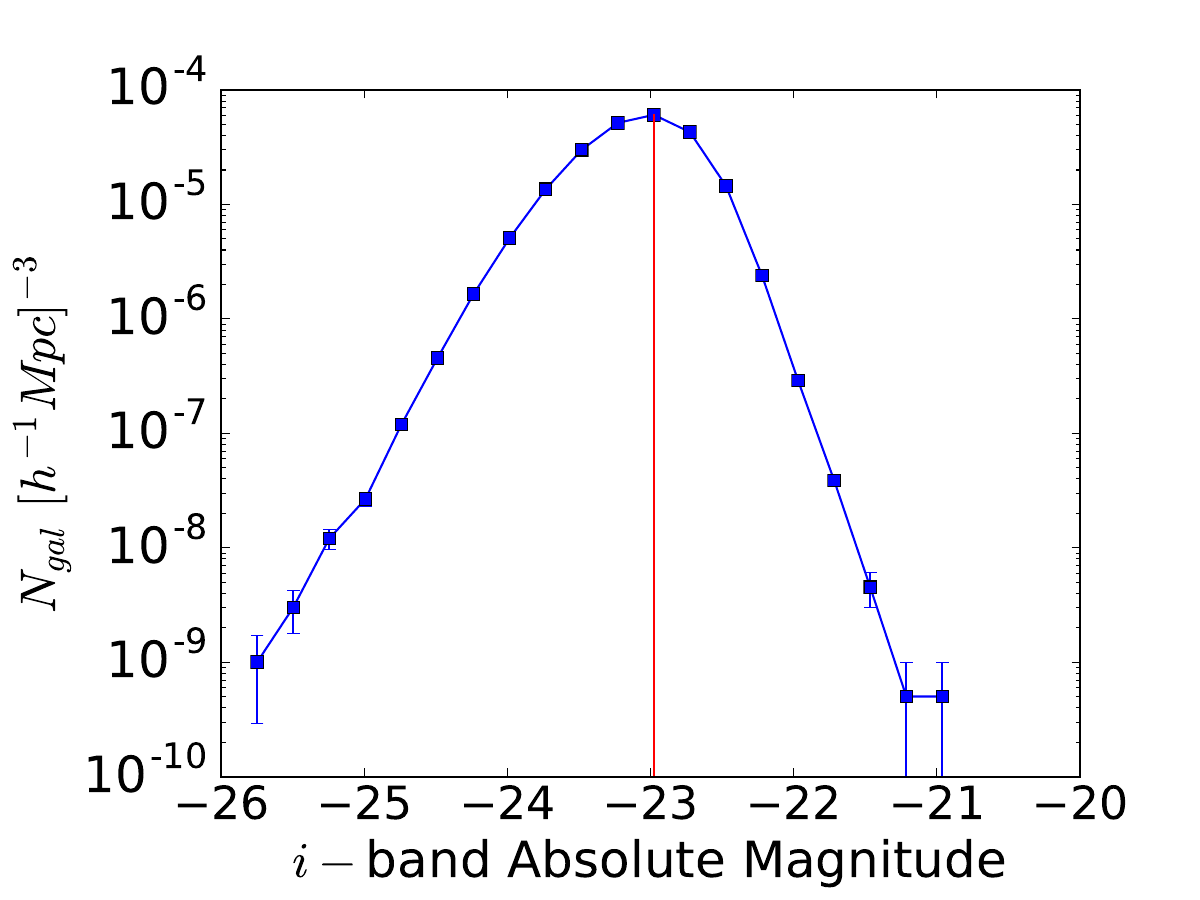}
\includegraphics[width=5.5cm, height=5.0cm]{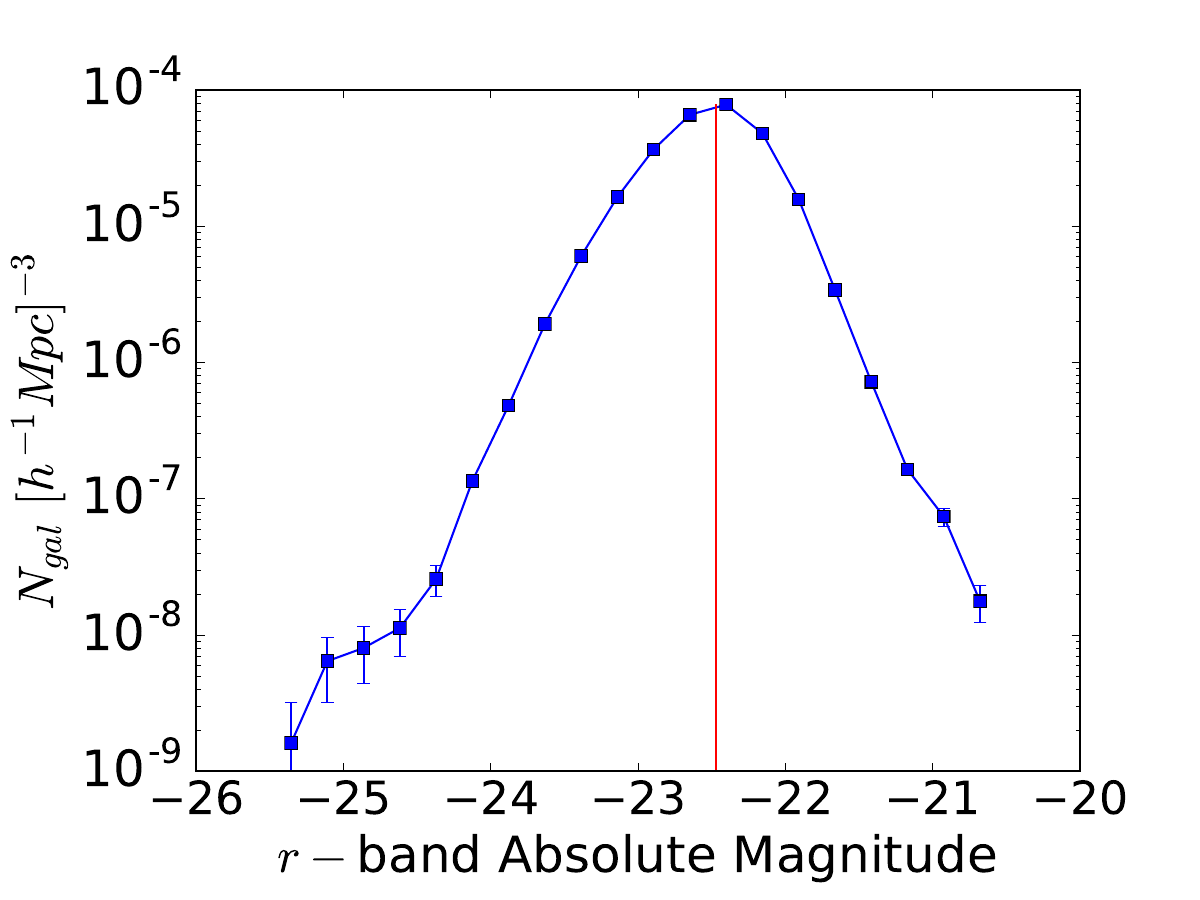}
\includegraphics[width=5.5cm, height=5.0cm]{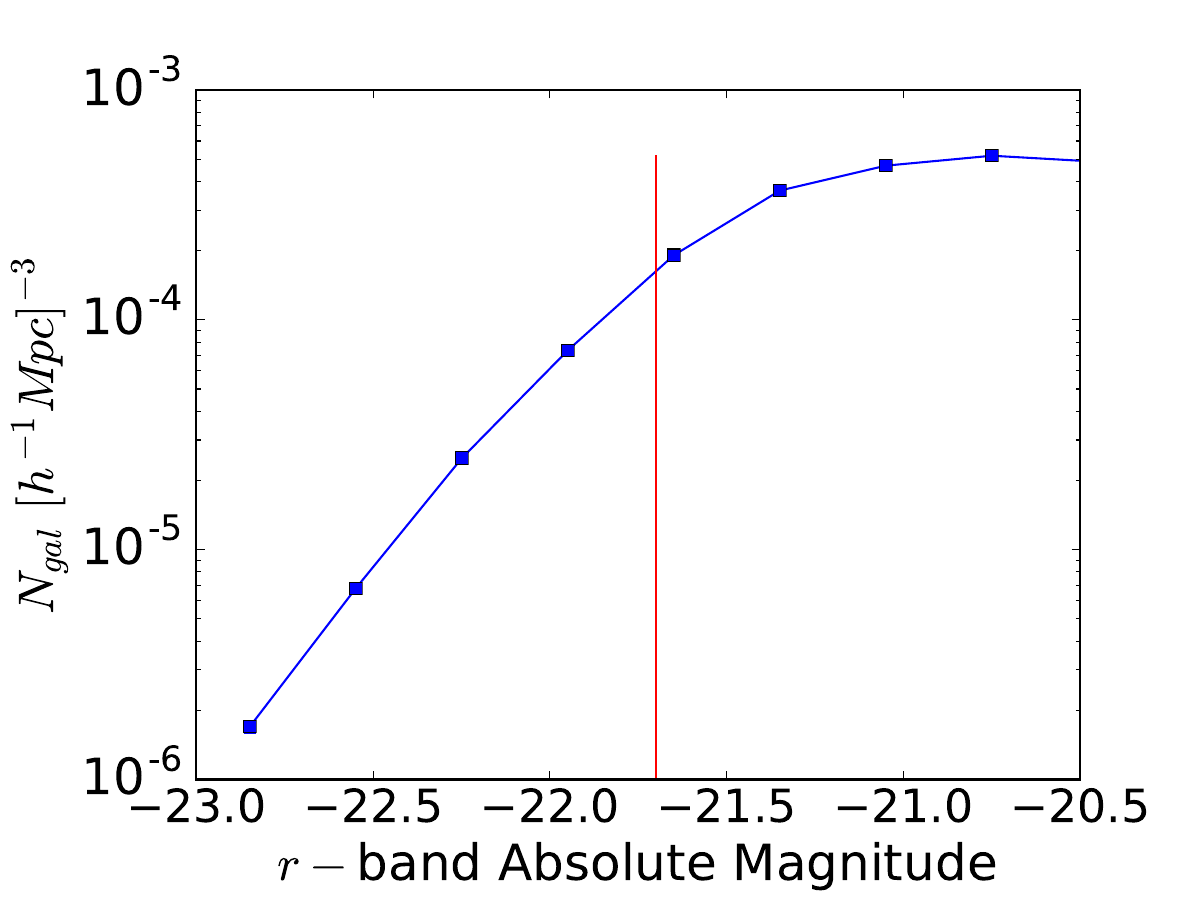}
\caption{$Top$ $panels:$ Points with errors show measurements of the projected correlation $w_{p}$ for the bright subsamples of CMASS (left), LOWZ (middle) and SDSS (right). In each panel, the number density of the sample is the same as the eBOSS+BOSS sample. Errors are obtained by jackknife sampling. The solid curves represent the best-fit HOD model to each sample. The dotted curves show the one-halo and two-halo terms. The fractional residuals of each fit are shown below. $Bottom$ $panels:$ The distribution of absolute magnitudes of the galaxies in the samples we used in the analysis. The errorbars are Poisson. The red vertical line corresponds to the cut we applied to each sample. All galaxies brighter than this line represent a sample with the same space density. These lines indicate that the CMASS and LOWZ subsamples are significantly more complete than the overall CMASS and LOWZ samples.}
\label{wp_sub}
\end{center}
\end{figure}

For the CMASS and LOWZ samples, we do the following. CMASS galaxies are first restricted to the redshift range $z=[0.47, 0.67]$. LOWZ galaxies are restricted to the range $z=[0.2, 0.4]$. Within these ranges, we rank order all galaxies by their absolute magnitude---$i$-band for CMASS and $r$-band for LOWZ. These choices of band correspond to the bands in which each sample was selected. The bottom panels in Figure \ref{wp_sub} show the distribution of absolute magnitudes in all three samples. For the SDSS-MGS, the sample is volume-limited up to $M_r-5\log h=-21.5$. For brighter magnitudes, the distribution of galaxy magnitudes is the same as the luminosity function of galaxies (ie., \citealt{Blanton_2005b}), but will deviate from the true luminosity function at fainter magnitudes. For the CMASS and LOWZ samples, the incompleteness of the samples is much more apparent. For both samples, the distribution of magnitudes more closely resembles a lognormal function. In each panel, the red line indicates the magnitude threshold utilized to create a sample with the number density of the eBOSS+BOSS sample. For the LOWZ and CMASS samples, this threshold lies at the peak of the magnitude distribution. Thus, our subsamples are more complete than the fiducial CMASS and LOWZ samples (\citealt{Reid_2016}). We note that we do not perform $k$-corrections for the LOWZ and CMASS samples. \cite{tinker_etal:16_boss} found that employing $k$-corrections on CMASS galaxies did not change the amplitude of their clustering, which is the quantity of interest for this analysis.

For each sample, we measure the projected correlation function. These data are shown in the upper panels of Figure \ref{wp_sub} as the points with error bars. Errors are calculated by jackknife sampling of the survey area into 25 subsamples. For each measurement of $w_p(r_p)$, we repeat the HOD analysis using the same HOD parameterization as used in the eBOSS+BOSS sample. The solid curves show the best-fit HOD for each sample. The fractional residuals of the fits are shown below the $w_p$ panels. We obtain the measurements of the bias of each sample from these HOD fits. These are the bias values used in \S 4.2 and Figure \ref{bias}.


\section{incompleteness in the HOD analysis}
\label{incomplet}
\setcounter{figure}{0}

\begin{figure}[htbp]
\begin{center}
\includegraphics[width=8.5cm, height=6.5cm]{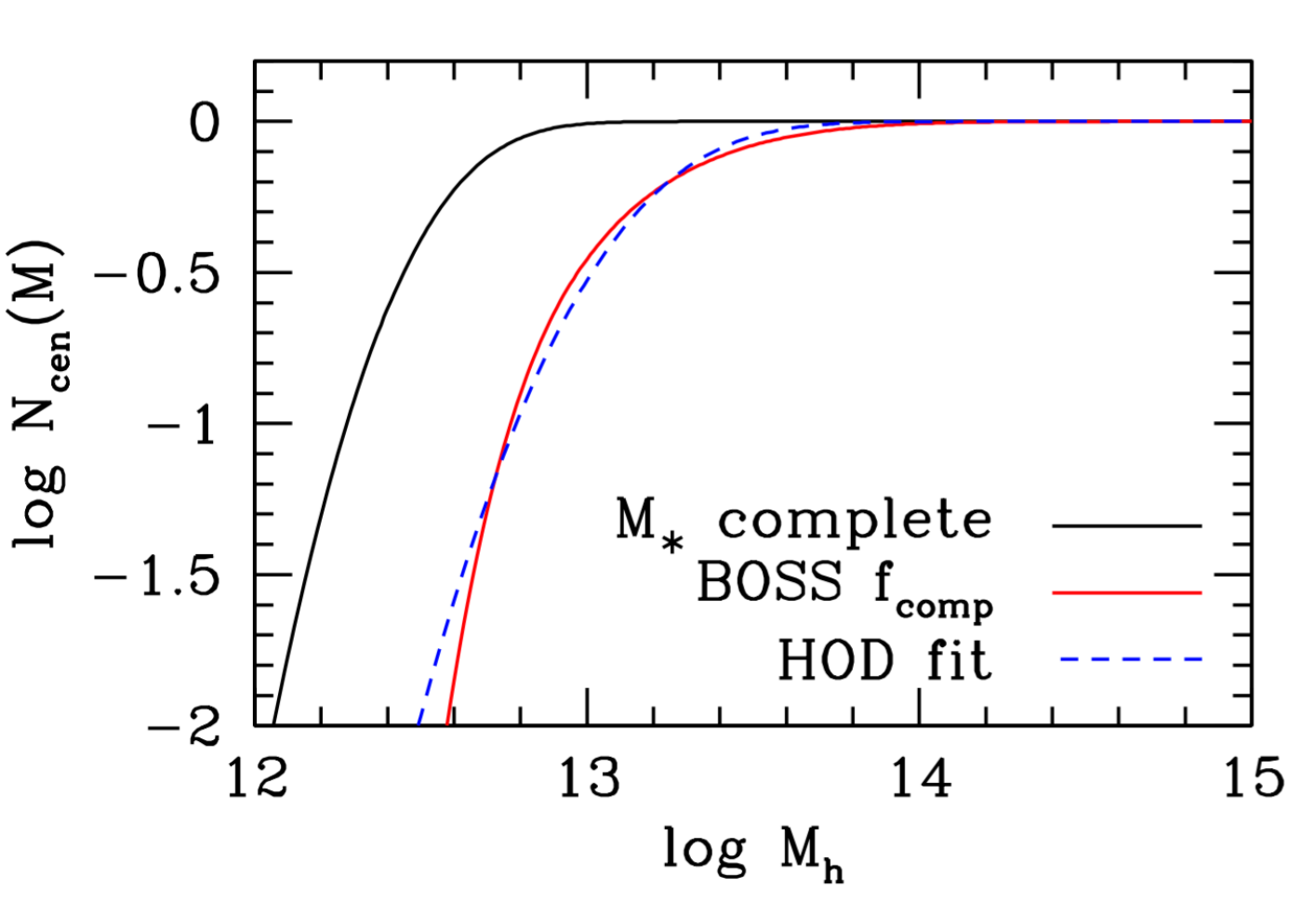}
\caption{The black solid curve shows the expected central occupation function for a sample that is complete for all galaxies more massive than $10^{11}$ $M_\odot$, using the stellar-to-halo mass relation of \cite{tinker_etal:16_boss}. The red solid line applies the stellar mass completeness of the BOSS CMASS sample found in \cite{tinker_etal:16_boss}. The dashed blue curve in a central occupation model, using Eq. (8), that matches both the number density and the bias of the red curve. }
\label{compl}
\end{center}
\end{figure}

In this appendix we demonstrate that the standard HOD approach is sufficient for analyzing incomplete samples such as LRGs. Although there is limited information about the completeness of the eBOSS sample, there are robust analyses of the completeness in BOSS (\citealt{tinker_etal:16_boss, leauthaud_etal:16}). Figure \ref{compl} shows the central occupation function for a sample of galaxies complete down to $\rm{M}_{\rm{gal}} > 10^{11} \rm{M}_{\odot}$, using the stellar-to-halo mass relation derived in \cite{tinker_etal:16_boss}. The solid red curve shows $\rm{N}_{\rm{cen}}$ after convolving the complete sample with the incompleteness function of BOSS CMASS galaxies found in \cite{tinker_etal:16_boss}, which is consistent with that found in \cite{leauthaud_etal:16} (cf., Figure 3 in \cite{tinker_etal:16_boss}). \cite{tinker_etal:16_boss} found that CMASS galaxies are 50\% complete at $\rm{M}_{\ast}=10^{11.4}$ $M_{\odot}$. After applying the incompleteness function, the number density of central galaxies is $2.63\times 10^{-4}$ $(h^{-1} \text{Mpc})^{-3}$, and the large-scale bias is 2.08. 

The dashed blue curve shows a central occupation function using Eq. (8). Although there are differences in the shape of the occupation function, the number density and bias of this function are $2.61\times 10^{-4}$ $(h^{-1} \text{Mpc})^{-3}$ and 2.09, respectively. Changing Eq. (8) to track the shape of red curve more exactly would not yield a substantive change in the clustering properties of the HOD itself. Thus, although the HOD derived in this analysis is only attributable to the eBOSS+BOSS sample itself, the parameterization used in this analysis is sufficient for modeling a sample with this type of incompleteness.

\bibliographystyle{apj}
\bibliography{LRGclustering}

\begin{thebibliography}{}
\expandafter\ifx\csname natexlab\endcsname\relax\def\natexlab#1{#1}\fi

\bibitem[{{Abazajian} {et~al.}(2004){Abazajian}, {Adelman-McCarthy},
  {Ag{\"u}eros}, {Allam}, {Anderson}, {Anderson}, {Annis}, {Bahcall}, {Baldry},
  {Bastian}, {Berlind}, {Bernardi}, {Blanton}, {Bochanski}, {Boroski},
  {Briggs}, {Brinkmann}, {Brunner}, {Budav{\'a}ri}, {Carey}, {Carliles},
  {Castander}, {Connolly}, {Csabai}, {Doi}, {Dong}, {Eisenstein}, {Evans},
  {Fan}, {Finkbeiner}, {Friedman}, {Frieman}, {Fukugita}, {Gal}, {Gillespie},
  {Glazebrook}, {Gray}, {Grebel}, {Gunn}, {Gurbani}, {Hall}, {Hamabe},
  {Harris}, {Harris}, {Harvanek}, {Heckman}, {Hendry}, {Hennessy}, {Hindsley},
  {Hogan}, {Hogg}, {Holmgren}, {Ichikawa}, {Ichikawa}, {Ivezi{\'c}}, {Jester},
  {Johnston}, {Jorgensen}, {Kent}, {Kleinman}, {Knapp}, {Kniazev}, {Kron},
  {Krzesinski}, {Kunszt}, {Kuropatkin}, {Lamb}, {Lampeitl}, {Lee}, {Leger},
  {Li}, {Lin}, {Loh}, {Long}, {Loveday}, {Lupton}, {Malik}, {Margon},
  {Matsubara}, {McGehee}, {McKay}, {Meiksin}, {Munn}, {Nakajima}, {Nash},
  {Neilsen}, {Newberg}, {Newman}, {Nichol}, {Nicinski}, {Nieto-Santisteban},
  {Nitta}, {Okamura}, {O'Mullane}, {Ostriker}, {Owen}, {Padmanabhan},
  {Peoples}, {Pier}, {Pope}, {Quinn}, {Richards}, {Richmond}, {Rix}, {Rockosi},
  {Schlegel}, {Schneider}, {Scranton}, {Sekiguchi}, {Seljak}, {Sergey},
  {Sesar}, {Sheldon}, {Shimasaku}, {Siegmund}, {Silvestri}, {Smith}, {Smol{\v
  c}i{\'c}}, {Snedden}, {Stebbins}, {Stoughton}, {Strauss}, {SubbaRao},
  {Szalay}, {Szapudi}, {Szkody}, {Szokoly}, {Tegmark}, {Teodoro}, {Thakar},
  {Tremonti}, {Tucker}, {Uomoto}, {Vanden Berk}, {Vandenberg}, {Vogeley},
  {Voges}, {Vogt}, {Walkowicz}, {Wang}, {Weinberg}, {West}, {White}, {Wilhite},
  {Xu}, {Yanny}, {Yasuda}, {Yip}, {Yocum}, {York}, {Zehavi}, {Zibetti}, \&
  {Zucker}}]{Abazajian_2004}
{Abazajian}, K., {Adelman-McCarthy}, J.~K., {Ag{\"u}eros}, M.~A., {et~al.}
  2004, \aj, 128, 502

\bibitem[{{Anderson} {et~al.}(2012){Anderson}, {Aubourg}, {Bailey}, {Bizyaev},
  {Blanton}, {Bolton}, {Brinkmann}, {Brownstein}, {Burden}, {Cuesta}, {da
  Costa}, {Dawson}, {de Putter}, {Eisenstein}, {Gunn}, {Guo}, {Hamilton},
  {Harding}, {Ho}, {Honscheid}, {Kazin}, {Kirkby}, {Kneib}, {Labatie},
  {Loomis}, {Lupton}, {Malanushenko}, {Malanushenko}, {Mandelbaum}, {Manera},
  {Maraston}, {McBride}, {Mehta}, {Mena}, {Montesano}, {Muna}, {Nichol},
  {Nuza}, {Olmstead}, {Oravetz}, {Padmanabhan}, {Palanque-Delabrouille}, {Pan},
  {Parejko}, {P{\^a}ris}, {Percival}, {Petitjean}, {Prada}, {Reid}, {Roe},
  {Ross}, {Ross}, {Samushia}, {S{\'a}nchez}, {Schlegel}, {Schneider},
  {Sc{\'o}ccola}, {Seo}, {Sheldon}, {Simmons}, {Skibba}, {Strauss}, {Swanson},
  {Thomas}, {Tinker}, {Tojeiro}, {Maga{\~n}a}, {Verde}, {Wagner}, {Wake},
  {Weaver}, {Weinberg}, {White}, {Xu}, {Y{\`e}che}, {Zehavi}, \&
  {Zhao}}]{Anderson_2012}
{Anderson}, L., {Aubourg}, E., {Bailey}, S., {et~al.} 2012, \mnras, 427, 3435

\bibitem[{{Behroozi} {et~al.}(2013{\natexlab{a}}){Behroozi}, {Marchesini},
  {Wechsler}, {Muzzin}, {Papovich}, \& {Stefanon}}]{Behroozi_2013c}
{Behroozi}, P.~S., {Marchesini}, D., {Wechsler}, R.~H., {et~al.}
  2013{\natexlab{a}}, \apjl, 777, L10

\bibitem[{{Behroozi} {et~al.}(2013{\natexlab{b}}){Behroozi}, {Wechsler}, \&
  {Wu}}]{Behroozi_2013b}
{Behroozi}, P.~S., {Wechsler}, R.~H., \& {Wu}, H.-Y. 2013{\natexlab{b}}, \apj,
  762, 109

\bibitem[{{Behroozi} {et~al.}(2013{\natexlab{c}}){Behroozi}, {Wechsler}, {Wu},
  {Busha}, {Klypin}, \& {Primack}}]{Behroozi_2013}
{Behroozi}, P.~S., {Wechsler}, R.~H., {Wu}, H.-Y., {et~al.} 2013{\natexlab{c}},
  \apj, 763, 18

\bibitem[{Benson {et~al.}(2000)Benson, Cole, Frenk, Baugh, \&
  Lacey}]{Benson_2000}
Benson, A.~J., Cole, S., Frenk, C.~S., Baugh, C.~M., \& Lacey, C.~G. 2000,
  Monthly Notices of the Royal Astronomical Society, 311, 793

\bibitem[{Berlind \& Weinberg(2002)}]{HOD_Weinberg}
Berlind, A.~A., \& Weinberg, D.~H. 2002, The Astrophysical Journal, 575, 587

\bibitem[{{Blanton} {et~al.}(2005{\natexlab{a}}){Blanton}, {Lupton},
  {Schlegel}, {Strauss}, {Brinkmann}, {Fukugita}, \& {Loveday}}]{Blanton_2005b}
{Blanton}, M.~R., {Lupton}, R.~H., {Schlegel}, D.~J., {et~al.}
  2005{\natexlab{a}}, \apj, 631, 208

\bibitem[{{Blanton} {et~al.}(2005{\natexlab{b}}){Blanton}, {Schlegel},
  {Strauss}, {Brinkmann}, {Finkbeiner}, {Fukugita}, {Gunn}, {Hogg},
  {Ivezi{\'c}}, {Knapp}, {Lupton}, {Munn}, {Schneider}, {Tegmark}, \&
  {Zehavi}}]{Blanton_2005a}
{Blanton}, M.~R., {Schlegel}, D.~J., {Strauss}, M.~A., {et~al.}
  2005{\natexlab{b}}, \aj, 129, 2562

\bibitem[{{Bolton} {et~al.}(2012){Bolton}, {Schlegel}, {Aubourg}, {Bailey},
  {Bhardwaj}, {Brownstein}, {Burles}, {Chen}, {Dawson}, {Eisenstein}, {Gunn},
  {Knapp}, {Loomis}, {Lupton}, {Maraston}, {Muna}, {Myers}, {Olmstead},
  {Padmanabhan}, {P{\^a}ris}, {Percival}, {Petitjean}, {Rockosi}, {Ross},
  {Schneider}, {Shu}, {Strauss}, {Thomas}, {Tremonti}, {Wake}, {Weaver}, \&
  {Wood-Vasey}}]{Bolton_2012}
{Bolton}, A.~S., {Schlegel}, D.~J., {Aubourg}, {\'E}., {et~al.} 2012, \aj, 144,
  144

\bibitem[{Cole {et~al.}(2005)Cole, Percival, Peacock, Norberg, Baugh, Frenk,
  Baldry, Bland-Hawthorn, Bridges, Cannon, Colless, Collins, Couch, Cross,
  Dalton, Eke, de~Propris, Driver, Efstathiou, Ellis, Glazebrook, Jackson,
  Jenkins, Lahav, Lewis, Lumsden, Maddox, Madgwick, Peterson, Sutherland,
  Taylor, \& 2dFGRS Team}]{Cole_2005}
Cole, S., Percival, W.~J., Peacock, J.~A., {et~al.} 2005, Monthly Notices of
  the Royal Astronomical Society, 362, 505

\bibitem[{{Contreras} {et~al.}(2017){Contreras}, {Zehavi}, {Baugh}, {Padilla},
  \& {Norberg}}]{Contreras_2017}
{Contreras}, S., {Zehavi}, I., {Baugh}, C.~M., {Padilla}, N., \& {Norberg}, P.
  2017, \mnras, 465, 2833

\bibitem[{{Cool} {et~al.}(2008){Cool}, {Eisenstein}, {Fan}, {Fukugita},
  {Jiang}, {Maraston}, {Meiksin}, {Schneider}, \& {Wake}}]{cool_etal:08}
{Cool}, R.~J., {Eisenstein}, D.~J., {Fan}, X., {et~al.} 2008, \apj, 682, 919

\bibitem[{{Cool} {et~al.}(2012){Cool}, {Eisenstein}, {Kochanek}, {Brown},
  {Caldwell}, {Dey}, {Forman}, {Hickox}, {Jannuzi}, {Jones}, {Moustakas}, \&
  {Murray}}]{Cool_2012}
{Cool}, R.~J., {Eisenstein}, D.~J., {Kochanek}, C.~S., {et~al.} 2012, \apj,
  748, 10

\bibitem[{Cooray \& Sheth(2002)}]{Cooray_2002}
Cooray, A., \& Sheth, R. 2002, Physics Reports, 372, 1

\bibitem[{{Davis} \& {Peebles}(1983)}]{Davis_1983}
{Davis}, M., \& {Peebles}, P.~J.~E. 1983, \apj, 267, 465

\bibitem[{Dawson {et~al.}(2013)Dawson, Schlegel, Ahn, Anderson, Éric Aubourg,
  Bailey, Barkhouser, Bautista, Beifiori, Berlind, Bhardwaj, Bizyaev, Blake,
  Blanton, Blomqvist, Bolton, Borde, Bovy, Brandt, Brewington, Brinkmann,
  Brown, Brownstein, Bundy, Busca, Carithers, Carnero, Carr, Chen, Comparat,
  Connolly, Cope, Croft, Cuesta, da~Costa, Davenport, Delubac, de~Putter,
  Dhital, Ealet, Ebelke, Eisenstein, Escoffier, Fan, Ak, Finley, Font-Ribera,
  Génova-Santos, Gunn, Guo, Haggard, Hall, Hamilton, Harris, Harris, Ho, Hogg,
  Holder, Honscheid, Huehnerhoff, Jordan, Jordan, Kauffmann, Kazin, Kirkby,
  Klaene, Kneib, Goff, Lee, Long, Loomis, Lundgren, Lupton, Maia, Makler,
  Malanushenko, Malanushenko, Mandelbaum, Manera, Maraston, Margala, Masters,
  McBride, McDonald, McGreer, McMahon, Mena, Miralda-Escudé, Montero-Dorta,
  Montesano, Muna, Myers, Naugle, Nichol, Noterdaeme, Nuza, Olmstead, Oravetz,
  Oravetz, Owen, Padmanabhan, Palanque-Delabrouille, Pan, Parejko, Pâris,
  Percival, Pérez-Fournon, Pérez-Ràfols, Petitjean, Pfaffenberger, Pforr,
  Pieri, Prada, Price-Whelan, Raddick, Rebolo, Rich, Richards, Rockosi, Roe,
  Ross, Ross, Rossi, Rubiño-Martin, Samushia, Sánchez, Sayres, Schmidt,
  Schneider, Scóccola, Seo, Shelden, Sheldon, Shen, Shu, Slosar, Smee,
  Snedden, Stauffer, Steele, Strauss, Streblyanska, Suzuki, Swanson, Tal,
  Tanaka, Thomas, Tinker, Tojeiro, Tremonti, Magaña, Verde, Viel, Wake,
  Watson, Weaver, Weinberg, Weiner, West, White, Wood-Vasey, Yeche, Zehavi,
  Zhao, \& Zheng}]{Dawson_BOSS}
Dawson, K.~S., Schlegel, D.~J., Ahn, C.~P., {et~al.} 2013, The Astronomical
  Journal, 145, 10

\bibitem[{{Dawson} {et~al.}(2016){Dawson}, {Kneib}, {Percival}, {Alam},
  {Albareti}, {Anderson}, {Armengaud}, {Aubourg}, {Bailey}, {Bautista},
  {Berlind}, {Bershady}, {Beutler}, {Bizyaev}, {Blanton}, {Blomqvist},
  {Bolton}, {Bovy}, {Brandt}, {Brinkmann}, {Brownstein}, {Burtin}, {Busca},
  {Cai}, {Chuang}, {Clerc}, {Comparat}, {Cope}, {Croft}, {Cruz-Gonzalez}, {da
  Costa}, {Cousinou}, {Darling}, {de la Macorra}, {de la Torre}, {Delubac}, {du
  Mas des Bourboux}, {Dwelly}, {Ealet}, {Eisenstein}, {Eracleous}, {Escoffier},
  {Fan}, {Finoguenov}, {Font-Ribera}, {Frinchaboy}, {Gaulme}, {Georgakakis},
  {Green}, {Guo}, {Guy}, {Ho}, {Holder}, {Huehnerhoff}, {Hutchinson}, {Jing},
  {Jullo}, {Kamble}, {Kinemuchi}, {Kirkby}, {Kitaura}, {Klaene}, {Laher},
  {Lang}, {Laurent}, {Le Goff}, {Li}, {Liang}, {Lima}, {Lin}, {Lin}, {Lin},
  {Long}, {Lundgren}, {MacDonald}, {Geimba Maia}, {Malanushenko},
  {Malanushenko}, {Mariappan}, {McBride}, {McGreer}, {M{\'e}nard}, {Merloni},
  {Meza}, {Montero-Dorta}, {Muna}, {Myers}, {Nandra}, {Naugle}, {Newman},
  {Noterdaeme}, {Nugent}, {Ogando}, {Olmstead}, {Oravetz}, {Oravetz},
  {Padmanabhan}, {Palanque-Delabrouille}, {Pan}, {Parejko}, {P{\^a}ris},
  {Peacock}, {Petitjean}, {Pieri}, {Pisani}, {Prada}, {Prakash}, {Raichoor},
  {Reid}, {Rich}, {Ridl}, {Rodriguez-Torres}, {Carnero Rosell}, {Ross},
  {Rossi}, {Ruan}, {Salvato}, {Sayres}, {Schneider}, {Schlegel}, {Seljak},
  {Seo}, {Sesar}, {Shandera}, {Shu}, {Slosar}, {Sobreira}, {Streblyanska},
  {Suzuki}, {Taylor}, {Tao}, {Tinker}, {Tojeiro}, {Vargas-Maga{\~n}a}, {Wang},
  {Weaver}, {Weinberg}, {White}, {Wood-Vasey}, {Yeche}, {Zhai}, {Zhao}, {Zhao},
  {Zheng}, {Ben Zhu}, \& {Zou}}]{eBOSS_Dawson}
{Dawson}, K.~S., {Kneib}, J.-P., {Percival}, W.~J., {et~al.} 2016, \aj, 151, 44

\bibitem[{Eisenstein {et~al.}(2005)Eisenstein, Zehavi, Hogg, Scoccimarro,
  Blanton, Nichol, Scranton, Seo, Tegmark, Zheng, Anderson, Annis, Bahcall,
  Brinkmann, Burles, Castander, Connolly, Csabai, Doi, Fukugita, Frieman,
  Glazebrook, Gunn, Hendry, Hennessy, Ivezić, Kent, Knapp, Lin, Loh, Lupton,
  Margon, McKay, Meiksin, Munn, Pope, Richmond, Schlegel, Schneider, Shimasaku,
  Stoughton, Strauss, SubbaRao, Szalay, Szapudi, Tucker, Yanny, \&
  York}]{Eisenstein_2005}
Eisenstein, D.~J., Zehavi, I., Hogg, D.~W., {et~al.} 2005, The Astrophysical
  Journal, 633, 560

\bibitem[{{Eisenstein} {et~al.}(2011){Eisenstein}, {Weinberg}, {Agol},
  {Aihara}, {Allende Prieto}, {Anderson}, {Arns}, {Aubourg}, {Bailey},
  {Balbinot}, \& et~al.}]{Eisenstein_2011}
{Eisenstein}, D.~J., {Weinberg}, D.~H., {Agol}, E., {et~al.} 2011, \aj, 142, 72

\bibitem[{Fisher {et~al.}(1994)Fisher, Davis, Strauss, Yahil, \&
  Huchra}]{Fisher_1994}
Fisher, K.~B., Davis, M., Strauss, M.~A., Yahil, A., \& Huchra, J. 1994,
  Monthly Notices of the Royal Astronomical Society, 266, 50

\bibitem[{Fry(1996)}]{Fry_bias}
Fry, J.~N. 1996, The Astrophysical Journal Letters, 461, L65

\bibitem[{{Fukugita} {et~al.}(1996){Fukugita}, {Ichikawa}, {Gunn}, {Doi},
  {Shimasaku}, \& {Schneider}}]{Fukugita_1996}
{Fukugita}, M., {Ichikawa}, T., {Gunn}, J.~E., {et~al.} 1996, \aj, 111, 1748

\bibitem[{{Gu} {et~al.}(2016){Gu}, {Conroy}, \& {Behroozi}}]{Gu_2016}
{Gu}, M., {Conroy}, C., \& {Behroozi}, P. 2016, ArXiv e-prints,
  arXiv:1602.01099

\bibitem[{{Gunn} {et~al.}(2006){Gunn}, {Siegmund}, {Mannery}, {Owen}, {Hull},
  {Leger}, {Carey}, {Knapp}, {York}, {Boroski}, {Kent}, {Lupton}, {Rockosi},
  {Evans}, {Waddell}, {Anderson}, {Annis}, {Barentine}, {Bartoszek}, {Bastian},
  {Bracker}, {Brewington}, {Briegel}, {Brinkmann}, {Brown}, {Carr},
  {Czarapata}, {Drennan}, {Dombeck}, {Federwitz}, {Gillespie}, {Gonzales},
  {Hansen}, {Harvanek}, {Hayes}, {Jordan}, {Kinney}, {Klaene}, {Kleinman},
  {Kron}, {Kresinski}, {Lee}, {Limmongkol}, {Lindenmeyer}, {Long}, {Loomis},
  {McGehee}, {Mantsch}, {Neilsen}, {Neswold}, {Newman}, {Nitta}, {Peoples},
  {Pier}, {Prieto}, {Prosapio}, {Rivetta}, {Schneider}, {Snedden}, \&
  {Wang}}]{APO_Gunn}
{Gunn}, J.~E., {Siegmund}, W.~A., {Mannery}, E.~J., {et~al.} 2006, \aj, 131,
  2332

\bibitem[{{Guo} {et~al.}(2013){Guo}, {Zehavi}, {Zheng}, {Weinberg}, {Berlind},
  {Blanton}, {Chen}, {Eisenstein}, {Ho}, {Kazin}, {Manera}, {Maraston},
  {McBride}, {Nuza}, {Padmanabhan}, {Parejko}, {Percival}, {Ross}, {Ross},
  {Samushia}, {S{\'a}nchez}, {Schlegel}, {Schneider}, {Skibba}, {Swanson},
  {Tinker}, {Tojeiro}, {Wake}, {White}, {Bahcall}, {Bizyaev}, {Brewington},
  {Bundy}, {da Costa}, {Ebelke}, {Malanushenko}, {Malanushenko}, {Oravetz},
  {Rossi}, {Simmons}, {Snedden}, {Streblyanska}, \& {Thomas}}]{Guo_2013}
{Guo}, H., {Zehavi}, I., {Zheng}, Z., {et~al.} 2013, \apj, 767, 122

\bibitem[{{Guo} {et~al.}(2014){Guo}, {Zheng}, {Zehavi}, {Xu}, {Eisenstein},
  {Weinberg}, {Bahcall}, {Berlind}, {Comparat}, {McBride}, {Ross}, {Schneider},
  {Skibba}, {Swanson}, {Tinker}, {Tojeiro}, \& {Wake}}]{Guo_2014}
{Guo}, H., {Zheng}, Z., {Zehavi}, I., {et~al.} 2014, \mnras, 441, 2398

\bibitem[{Hawkins {et~al.}(2003)Hawkins, Maddox, Cole, Lahav, Madgwick,
  Norberg, Peacock, Baldry, Baugh, Bland-Hawthorn, Bridges, Cannon, Colless,
  Collins, Couch, Dalton, de~Propris, Driver, Efstathiou, Ellis, Frenk,
  Glazebrook, Jackson, Jones, Lewis, Lumsden, Percival, Peterson, Sutherland,
  \& Taylor}]{Hawkins_2003}
Hawkins, E., Maddox, S., Cole, S., {et~al.} 2003, Monthly Notices of the Royal
  Astronomical Society, 346, 78

\bibitem[{{Hutchinson} {et~al.}(2016)}]{Hutchinson_16_redmonster}
{Hutchinson}, T., {et~al.} 2016, in preparation

\bibitem[{{Kravtsov} {et~al.}(2014){Kravtsov}, {Vikhlinin}, \&
  {Meshscheryakov}}]{kravtsov_etal:14}
{Kravtsov}, A., {Vikhlinin}, A., \& {Meshscheryakov}, A. 2014, \apj, submitted,
  arXiv:1401.7329, arXiv:1401.7329

\bibitem[{{Landy} \& {Szalay}(1993)}]{LS_1993}
{Landy}, S.~D., \& {Szalay}, A.~S. 1993, \apj, 412, 64

\bibitem[{{Leauthaud} {et~al.}(2016){Leauthaud}, {Bundy}, {Saito}, {Tinker},
  {Maraston}, {Tojeiro}, {Huang}, {Brownstein}, {Schneider}, \&
  {Thomas}}]{leauthaud_etal:16}
{Leauthaud}, A., {Bundy}, K., {Saito}, S., {et~al.} 2016, \mnras, 457, 4021

\bibitem[{{Lehmann} {et~al.}(2015){Lehmann}, {Mao}, {Becker}, {Skillman}, \&
  {Wechsler}}]{lehmann_etal:15}
{Lehmann}, B.~V., {Mao}, Y.-Y., {Becker}, M.~R., {Skillman}, S.~W., \&
  {Wechsler}, R.~H. 2015, ArXiv e-prints, arXiv:1510.05651

\bibitem[{{Li} {et~al.}(2012){Li}, {Jing}, {Mao}, {Han}, {Peng}, {Yang}, {Mo},
  \& {van den Bosch}}]{li_etal:12}
{Li}, C., {Jing}, Y.~P., {Mao}, S., {et~al.} 2012, \apj, 758, 50

\bibitem[{{Macci{\`o}} {et~al.}(2008){Macci{\`o}}, {Dutton}, \& {van den
  Bosch}}]{Maccio_2008}
{Macci{\`o}}, A.~V., {Dutton}, A.~A., \& {van den Bosch}, F.~C. 2008, \mnras,
  391, 1940

\bibitem[{{More} {et~al.}(2009){More}, {van den Bosch}, {Cacciato}, {Mo},
  {Yang}, \& {Li}}]{more_etal:09}
{More}, S., {van den Bosch}, F.~C., {Cacciato}, M., {et~al.} 2009, \mnras, 392,
  801

\bibitem[{{Moster} {et~al.}(2013){Moster}, {Naab}, \& {White}}]{moster_etal:13}
{Moster}, B.~P., {Naab}, T., \& {White}, S.~D.~M. 2013, \mnras, 428, 3121

\bibitem[{{Navarro} {et~al.}(1996){Navarro}, {Frenk}, \& {White}}]{NFW_1996}
{Navarro}, J.~F., {Frenk}, C.~S., \& {White}, S.~D.~M. 1996, \apj, 462, 563

\bibitem[{Norberg {et~al.}(2009)Norberg, Baugh, Gaztañaga, \&
  Croton}]{Norberg_2009}
Norberg, P., Baugh, C.~M., Gaztañaga, E., \& Croton, D.~J. 2009, Monthly
  Notices of the Royal Astronomical Society, 396, 19

\bibitem[{Parejko {et~al.}(2013)Parejko, Sunayama, Padmanabhan, Wake, Berlind,
  Bizyaev, Blanton, Bolton, van~den Bosch, Brinkmann, Brownstein, da~Costa,
  Eisenstein, Guo, Kazin, Maia, Malanushenko, Maraston, McBride, Nichol,
  Oravetz, Pan, Percival, Prada, Ross, Ross, Schlegel, Schneider, Simmons,
  Skibba, Tinker, Tojeiro, Weaver, Wetzel, White, Weinberg, Thomas, Zehavi, \&
  Zheng}]{Parejko_LOWZ}
Parejko, J.~K., Sunayama, T., Padmanabhan, N., {et~al.} 2013, Monthly Notices
  of the Royal Astronomical Society, 429, 98

\bibitem[{Peacock \& Smith(2000)}]{Peacock_2000}
Peacock, J.~A., \& Smith, R.~E. 2000, Monthly Notices of the Royal Astronomical
  Society, 318, 1144

\bibitem[{Peebles(1980)}]{Peebles1980large}
Peebles, P. J.~E. 1980, The large-scale structure of the universe (Princeton
  university press)

\bibitem[{{Prakash} {et~al.}(2016){Prakash}, {Licquia}, {Newman}, {Ross},
  {Myers}, {Dawson}, {Kneib}, {Percival}, {Bautista}, {Comparat}, {Tinker},
  {Schlegel}, {Tojeiro}, {Ho}, {Lang}, {Rao}, {McBride}, {Ben Zhu},
  {Brownstein}, {Bailey}, {Bolton}, {Delubac}, {Mariappan}, {Blanton}, {Reid},
  {Schneider}, {Seo}, {Carnero Rosell}, \& {Prada}}]{LRG_Prakash}
{Prakash}, A., {Licquia}, T.~C., {Newman}, J.~A., {et~al.} 2016, \apjs, 224, 34

\bibitem[{{Reddick} {et~al.}(2013){Reddick}, {Wechsler}, {Tinker}, \&
  {Behroozi}}]{reddick_etal:13}
{Reddick}, R.~M., {Wechsler}, R.~H., {Tinker}, J.~L., \& {Behroozi}, P.~S.
  2013, \apj, 771, 30

\bibitem[{{Reid} {et~al.}(2016){Reid}, {Ho}, {Padmanabhan}, {Percival},
  {Tinker}, {Tojeiro}, {White}, {Eisenstein}, {Maraston}, {Ross},
  {S{\'a}nchez}, {Schlegel}, {Sheldon}, {Strauss}, {Thomas}, {Wake}, {Beutler},
  {Bizyaev}, {Bolton}, {Brownstein}, {Chuang}, {Dawson}, {Harding}, {Kitaura},
  {Leauthaud}, {Masters}, {McBride}, {More}, {Olmstead}, {Oravetz}, {Nuza},
  {Pan}, {Parejko}, {Pforr}, {Prada}, {Rodr{\'{\i}}guez-Torres},
  {Salazar-Albornoz}, {Samushia}, {Schneider}, {Sc{\'o}ccola}, {Simmons}, \&
  {Vargas-Magana}}]{Reid_2016}
{Reid}, B., {Ho}, S., {Padmanabhan}, N., {et~al.} 2016, \mnras, 455, 1553

\bibitem[{{Reid} {et~al.}(2014){Reid}, {Seo}, {Leauthaud}, {Tinker}, \&
  {White}}]{Reid_2014}
{Reid}, B.~A., {Seo}, H.-J., {Leauthaud}, A., {Tinker}, J.~L., \& {White}, M.
  2014, \mnras, 444, 476

\bibitem[{{Riebe} {et~al.}(2011){Riebe}, {Partl}, {Enke}, {Forero-Romero},
  {Gottloeber}, {Klypin}, {Lemson}, {Prada}, {Primack}, {Steinmetz}, \&
  {Turchaninov}}]{Riebe_2011}
{Riebe}, K., {Partl}, A.~M., {Enke}, H., {et~al.} 2011, ArXiv e-prints,
  arXiv:1109.0003

\bibitem[{Seljak(2000)}]{Seljak_2000}
Seljak, U. 2000, Monthly Notices of the Royal Astronomical Society, 318, 203

\bibitem[{{Smee} {et~al.}(2013){Smee}, {Gunn}, {Uomoto}, {Roe}, {Schlegel},
  {Rockosi}, {Carr}, {Leger}, {Dawson}, {Olmstead}, {Brinkmann}, {Owen},
  {Barkhouser}, {Honscheid}, {Harding}, {Long}, {Lupton}, {Loomis}, {Anderson},
  {Annis}, {Bernardi}, {Bhardwaj}, {Bizyaev}, {Bolton}, {Brewington}, {Briggs},
  {Burles}, {Burns}, {Castander}, {Connolly}, {Davenport}, {Ebelke}, {Epps},
  {Feldman}, {Friedman}, {Frieman}, {Heckman}, {Hull}, {Knapp}, {Lawrence},
  {Loveday}, {Mannery}, {Malanushenko}, {Malanushenko}, {Merrelli}, {Muna},
  {Newman}, {Nichol}, {Oravetz}, {Pan}, {Pope}, {Ricketts}, {Shelden},
  {Sandford}, {Siegmund}, {Simmons}, {Smith}, {Snedden}, {Schneider},
  {SubbaRao}, {Tremonti}, {Waddell}, \& {York}}]{Smee_2013}
{Smee}, S.~A., {Gunn}, J.~E., {Uomoto}, A., {et~al.} 2013, \aj, 146, 32

\bibitem[{{Stoughton} {et~al.}(2002){Stoughton}, {Lupton}, {Bernardi},
  {Blanton}, {Burles}, {Castander}, {Connolly}, {Eisenstein}, {Frieman},
  {Hennessy}, {Hindsley}, {Ivezi{\'c}}, {Kent}, {Kunszt}, {Lee}, {Meiksin},
  {Munn}, {Newberg}, {Nichol}, {Nicinski}, {Pier}, {Richards}, {Richmond},
  {Schlegel}, {Smith}, {Strauss}, {SubbaRao}, {Szalay}, {Thakar}, {Tucker},
  {Vanden Berk}, {Yanny}, {Adelman}, {Anderson}, {Anderson}, {Annis},
  {Bahcall}, {Bakken}, {Bartelmann}, {Bastian}, {Bauer}, {Berman},
  {B{\"o}hringer}, {Boroski}, {Bracker}, {Briegel}, {Briggs}, {Brinkmann},
  {Brunner}, {Carey}, {Carr}, {Chen}, {Christian}, {Colestock}, {Crocker},
  {Csabai}, {Czarapata}, {Dalcanton}, {Davidsen}, {Davis}, {Dehnen},
  {Dodelson}, {Doi}, {Dombeck}, {Donahue}, {Ellman}, {Elms}, {Evans}, {Eyer},
  {Fan}, {Federwitz}, {Friedman}, {Fukugita}, {Gal}, {Gillespie}, {Glazebrook},
  {Gray}, {Grebel}, {Greenawalt}, {Greene}, {Gunn}, {de Haas}, {Haiman},
  {Haldeman}, {Hall}, {Hamabe}, {Hansen}, {Harris}, {Harris}, {Harvanek},
  {Hawley}, {Hayes}, {Heckman}, {Helmi}, {Henden}, {Hogan}, {Hogg}, {Holmgren},
  {Holtzman}, {Huang}, {Hull}, {Ichikawa}, {Ichikawa}, {Johnston}, {Kauffmann},
  {Kim}, {Kimball}, {Kinney}, {Klaene}, {Kleinman}, {Klypin}, {Knapp},
  {Korienek}, {Krolik}, {Kron}, {Krzesi{\'n}ski}, {Lamb}, {Leger},
  {Limmongkol}, {Lindenmeyer}, {Long}, {Loomis}, {Loveday}, {MacKinnon},
  {Mannery}, {Mantsch}, {Margon}, {McGehee}, {McKay}, {McLean}, {Menou},
  {Merelli}, {Mo}, {Monet}, {Nakamura}, {Narayanan}, {Nash}, {Neilsen},
  {Newman}, {Nitta}, {Odenkirchen}, {Okada}, {Okamura}, {Ostriker}, {Owen},
  {Pauls}, {Peoples}, {Peterson}, {Petravick}, {Pope}, {Pordes}, {Postman},
  {Prosapio}, {Quinn}, {Rechenmacher}, {Rivetta}, {Rix}, {Rockosi}, {Rosner},
  {Ruthmansdorfer}, {Sandford}, {Schneider}, {Scranton}, {Sekiguchi}, {Sergey},
  {Sheth}, {Shimasaku}, {Smee}, {Snedden}, {Stebbins}, {Stubbs}, {Szapudi},
  {Szkody}, {Szokoly}, {Tabachnik}, {Tsvetanov}, {Uomoto}, {Vogeley}, {Voges},
  {Waddell}, {Walterbos}, {Wang}, {Watanabe}, {Weinberg}, {White}, {White},
  {Wilhite}, {Wolfe}, {Yasuda}, {York}, {Zehavi}, \& {Zheng}}]{Stoughton_2002}
{Stoughton}, C., {Lupton}, R.~H., {Bernardi}, M., {et~al.} 2002, \aj, 123, 485

\bibitem[{Swanson {et~al.}(2008)Swanson, Tegmark, Hamilton, \&
  Hill}]{Swanson_Mangle}
Swanson, M. E.~C., Tegmark, M., Hamilton, A. J.~S., \& Hill, J.~C. 2008,
  Monthly Notices of the Royal Astronomical Society, 387, 1391

\bibitem[{{Tinker} {et~al.}(2008){Tinker}, {Kravtsov}, {Klypin}, {Abazajian},
  {Warren}, {Yepes}, {Gottl{\"o}ber}, \& {Holz}}]{Tinker_2008}
{Tinker}, J., {Kravtsov}, A.~V., {Klypin}, A., {et~al.} 2008, \apj, 688, 709

\bibitem[{{Tinker} {et~al.}(2010){Tinker}, {Robertson}, {Kravtsov}, {Klypin},
  {Warren}, {Yepes}, \& {Gottl{\"o}ber}}]{Tinker_2010}
{Tinker}, J.~L., {Robertson}, B.~E., {Kravtsov}, A.~V., {et~al.} 2010, \apj,
  724, 878

\bibitem[{Tinker {et~al.}(2005)Tinker, Weinberg, Zheng, \&
  Zehavi}]{Tinker_analytical}
Tinker, J.~L., Weinberg, D.~H., Zheng, Z., \& Zehavi, I. 2005, The
  Astrophysical Journal, 631, 41

\bibitem[{{Tinker} {et~al.}(2012){Tinker}, {Sheldon}, {Wechsler}, {Becker},
  {Rozo}, {Zu}, {Weinberg}, {Zehavi}, {Blanton}, {Busha}, \&
  {Koester}}]{Tinker_2012}
{Tinker}, J.~L., {Sheldon}, E.~S., {Wechsler}, R.~H., {et~al.} 2012, \apj, 745,
  16

\bibitem[{{Tinker} {et~al.}(2017){Tinker}, {Brownstein}, {Guo}, {Leauthaud},
  {Maraston}, {Masters}, {Montero-Dorta}, {Thomas}, {Tojeiro}, {Weiner},
  {Zehavi}, \& {Olmstead}}]{tinker_etal:16_boss}
{Tinker}, J.~L., {Brownstein}, J.~R., {Guo}, H., {et~al.} 2017, \apj, 839, 121

\bibitem[{{Wake} {et~al.}(2008){Wake}, {Sheth}, {Nichol}, {Baugh},
  {Bland-Hawthorn}, {Colless}, {Couch}, {Croom}, {de Propris}, {Drinkwater},
  {Edge}, {Loveday}, {Lam}, {Pimbblet}, {Roseboom}, {Ross}, {Schneider},
  {Shanks}, \& {Sharp}}]{wake_etal:08}
{Wake}, D.~A., {Sheth}, R.~K., {Nichol}, R.~C., {et~al.} 2008, \mnras, 387,
  1045

\bibitem[{{Wetzel} \& {White}(2010)}]{Wetzel_Martin_2010}
{Wetzel}, A.~R., \& {White}, M. 2010, \mnras, 403, 1072

\bibitem[{White {et~al.}(2001)White, Hernquist, \& Springel}]{Martin_2001}
White, M., Hernquist, L., \& Springel, V. 2001, The Astrophysical Journal
  Letters, 550, L129

\bibitem[{White {et~al.}(2014)White, Tinker, \& McBride}]{QPM_White}
White, M., Tinker, J.~L., \& McBride, C.~K. 2014, Monthly Notices of the Royal
  Astronomical Society, 437, 2594

\bibitem[{White {et~al.}(2011)White, Blanton, Bolton, Schlegel, Tinker,
  Berlind, da~Costa, Kazin, Lin, Maia, McBride, Padmanabhan, Parejko, Percival,
  Prada, Ramos, Sheldon, de~Simoni, Skibba, Thomas, Wake, Zehavi, Zheng,
  Nichol, Schneider, Strauss, Weaver, \& Weinberg}]{CMASS_Martin}
White, M., Blanton, M., Bolton, A., {et~al.} 2011, The Astrophysical Journal,
  728, 126

\bibitem[{{Wright} {et~al.}(2010){Wright}, {Eisenhardt}, {Mainzer}, {Ressler},
  {Cutri}, {Jarrett}, {Kirkpatrick}, {Padgett}, {McMillan}, {Skrutskie},
  {Stanford}, {Cohen}, {Walker}, {Mather}, {Leisawitz}, {Gautier}, {McLean},
  {Benford}, {Lonsdale}, {Blain}, {Mendez}, {Irace}, {Duval}, {Liu}, {Royer},
  {Heinrichsen}, {Howard}, {Shannon}, {Kendall}, {Walsh}, {Larsen}, {Cardon},
  {Schick}, {Schwalm}, {Abid}, {Fabinsky}, {Naes}, \& {Tsai}}]{WISE_Wright}
{Wright}, E.~L., {Eisenhardt}, P.~R.~M., {Mainzer}, A.~K., {et~al.} 2010, \aj,
  140, 1868

\bibitem[{Zehavi {et~al.}(2002)Zehavi, Blanton, Frieman, Weinberg, Mo, Strauss,
  Anderson, Annis, Bahcall, Bernardi, Briggs, Brinkmann, Burles, Carey,
  Castander, Connolly, Csabai, Dalcanton, Dodelson, Doi, Eisenstein, Evans,
  Finkbeiner, Friedman, Fukugita, Gunn, Hennessy, Hindsley, Željko Ivezić,
  Kent, Knapp, Kron, Kunszt, Lamb, Leger, Long, Loveday, Lupton, McKay,
  Meiksin, Merrelli, Munn, Narayanan, Newcomb, Nichol, Owen, Peoples, Pope,
  Rockosi, Schlegel, Schneider, Scoccimarro, Sheth, Siegmund, Smee, Snir,
  Stebbins, Stoughton, SubbaRao, Szalay, Szapudi, Tegmark, Tucker, Uomoto,
  Berk, Vogeley, Waddell, Yanny, York, \& the
  SDSS~Collaboration}]{Zehavi_nearest}
Zehavi, I., Blanton, M.~R., Frieman, J.~A., {et~al.} 2002, The Astrophysical
  Journal, 571, 172

\bibitem[{Zehavi {et~al.}(2005)Zehavi, Zheng, Weinberg, Frieman, Berlind,
  Blanton, Scoccimarro, Sheth, Strauss, Kayo, Suto, Fukugita, Nakamura,
  Bahcall, Brinkmann, Gunn, Hennessy, Željko Ivezić, Knapp, Loveday, Meiksin,
  Schlegel, Schneider, Szapudi, Tegmark, Vogeley, York, \& the
  SDSS~Collaboration}]{Zehavi_2005}
Zehavi, I., Zheng, Z., Weinberg, D.~H., {et~al.} 2005, The Astrophysical
  Journal, 630, 1

\bibitem[{{Zehavi} {et~al.}(2011){Zehavi}, {Zheng}, {Weinberg}, {Blanton},
  {Bahcall}, {Berlind}, {Brinkmann}, {Frieman}, {Gunn}, {Lupton}, {Nichol},
  {Percival}, {Schneider}, {Skibba}, {Strauss}, {Tegmark}, \&
  {York}}]{Zehavi_2011}
{Zehavi}, I., {Zheng}, Z., {Weinberg}, D.~H., {et~al.} 2011, \apj, 736, 59

\bibitem[{{Zhao} {et~al.}(2016){Zhao}, {Wang}, {Ross}, {Shandera}, {Percival},
  {Dawson}, {Kneib}, {Myers}, {Brownstein}, {Comparat}, {Delubac}, {Gao},
  {Hojjati}, {Koyama}, {McBride}, {Meza}, {Newman}, {Palanque-Delabrouille},
  {Pogosian}, {Prada}, {Rossi}, {Schneider}, {Seo}, {Tao}, {Wang}, {Y{\`e}che},
  {Zhang}, {Zhang}, {Zhou}, {Zhu}, \& {Zou}}]{Zhao_eBOSS}
{Zhao}, G.-B., {Wang}, Y., {Ross}, A.~J., {et~al.} 2016, \mnras, 457, 2377

\bibitem[{Zheng {et~al.}(2007)Zheng, Coil, \& Zehavi}]{Zheng_DEEP2}
Zheng, Z., Coil, A.~L., \& Zehavi, I. 2007, The Astrophysical Journal, 667, 760

\bibitem[{{Zheng} {et~al.}(2009){Zheng}, {Zehavi}, {Eisenstein}, {Weinberg}, \&
  {Jing}}]{Zheng_2009}
{Zheng}, Z., {Zehavi}, I., {Eisenstein}, D.~J., {Weinberg}, D.~H., \& {Jing},
  Y.~P. 2009, \apj, 707, 554

\bibitem[{{Zu} \& {Mandelbaum}(2016)}]{zu_mandelbaum:16}
{Zu}, Y., \& {Mandelbaum}, R. 2016, \mnras, 457, 4360

\end{thebibliography}

\end{document}